\documentclass[journal]{IEEEtran}

\usepackage[T1]{fontenc}
\usepackage[utf8]{inputenc}
\usepackage{cite}
\usepackage{graphicx}
\usepackage{textcomp}
\usepackage{url}
\usepackage{amsmath,amsthm,amssymb,amsfonts,bm,mathtools}
\usepackage{siunitx}
\usepackage{booktabs}
\usepackage{array}
\usepackage{placeins}
\usepackage[hidelinks]{hyperref}

\newcolumntype{L}[1]{>{\raggedright\arraybackslash}p{#1}}

\title{Automated Full-Sphere Measurement Methodology for
Reconfigurable Intelligent Surface Beam Steering in an Anechoic Chamber}

\author{%
Tobias Kancz, Robert Langwieser, and Philipp Svoboda\\[-0.2em]
{\small Christian Doppler Laboratory for Digital Twin Assisted AI for
Sustainable Radio Access Networks}\\[-0.1em]
{\small Institute of Telecommunications, TU Wien, 1040 Vienna, Austria}\\[-0.1em]
{\small Corresponding author: Tobias Kancz
(e-mail: tobias.kancz@tuwien.ac.at)}
}

\begin{document}
\maketitle

\begin{abstract}
Reconfigurable intelligent surfaces (RISs) need measurement methodologies that quantify the three-dimensional scattering response of a programmed surface, not only the received power of one bistatic link. This paper presents an automated anechoic-chamber methodology for full-sphere characterization of a varactor-based RIS around \SI{5}{GHz}. The developed platform combines a state-of-the-art spherical near-field scanner with a near-field-to-far-field transformation, which is suitable for antenna measurements, with a common, low-scattering mount for the RIS and the excitation antenna. It enables  programming of several independent DC voltages for remote RIS beam steering configuration via a control unit integrated into the chamber, which features a multichannel digital-to-analog converter.
The RIS states are synthesized from a pre-measured phase-voltage characteristic and evaluated against an unbiased-RIS reference using local angular power averages. Measurements at \SI{5.3}{GHz} serve as a proof of operation: the workflow programs one-dimensional and two-dimensional anomalous-reflection states, localizes the resulting beams over the sphere, and extracts target-window gain, maximum-window gain, and pointing error from the same dataset. For five measured RIS beam-steering states, the prescribed target windows show relative beam-steering gains up to \SI{8.8}{dB}, while the locally detected main-lobe windows show relative gains up to \SI{14.0}{dB} with an average of \SI{11.3}{dB}. The reported gain is not active amplification by the RIS; it is a reference-normalized local power ratio for a fixed measurement configuration.
\end{abstract}


\section{Introduction}
\label{sec:introduction}
\IEEEPARstart{R}{econfigurable} intelligent surfaces (RISs) are being investigated as a means of controlling the radio environment by imposing programmable phase, amplitude, or polarization responses on incident electromagnetic waves. Their appeal is strongest in coverage-extension and spatial-selectivity problems, where passive or nearly passive apertures can redirect energy without the power consumption and latency profile of an active relay \cite{Alsabah2021ACCESS,DiRenzo2020OJCOMS,Liu2021COMST,Basar2019ACCESS,Wu2020MCOM}. Recent surveys, tutorials, and analytical studies have clarified the communication-theoretic potential of RIS-aided links, the role of hardware constraints, and the need for electromagnetic-compliant models \cite{DiRenzo2020JSAC,Wu2021TCOMM,Hou2022TVT,Basar2024VTM}.

The experimental characterization of RIS hardware has not matured at the same rate. Existing campaigns typically emphasize one of three objectives: path-loss validation in fixed bistatic geometries \cite{Tang2021TWC}, channel-state datasets over selected angular or spatial grids \cite{Tewes2023VTC}, or system-level proof-of-concept measurements in chamber and outdoor scenarios \cite{Trichopoulos2022OJCOMS}. A recent survey of RIS parameters and measurement techniques also shows that reported results depend strongly on the chosen reference, angular scan, calibration plane, and bias sequence \cite{Rana2022Micromachines}. The same issue is reflected in ongoing RIS test-standardization work, which defines measurable RIS characteristics but still leaves the laboratory realization to the test facility \cite{ETSI2026RIS008}. In particular, planar angular cuts can miss off-plane sidelobes and can overestimate or underestimate the main beam when the true maximum is displaced from the measured cut. Manual changes of RIS state or geometry also introduce cable motion, thermal drift, and alignment error that are difficult to reproduce.

 Recent RIS prototypes have demonstrated continuous or multi-level phase control and reflected beamforming at sub-6 GHz, at \SI{9.6}{GHz}, and in the Ka band \cite{Ratajczak2024AnnTel,Chen2025ACCESS,Cifuentes2025OJCOMS,Wolff2023JAP}. These works establish viable hardware implementations, but most reported measurement evidence is still tied to selected receiver positions, selected angular cuts, or prototype-specific reference definitions. The missing step is a reproducible full-sphere characterization of the programmed scattering response for each RIS state.

Near-field-to-far-field (NF--FF) transformation is a mature part of antenna metrology, including spherical near-field scanning \cite{Yaghjian1986TAP,Hansen1988SNF,IEEEStd1492021}. Its direct use for RIS evaluation nevertheless needs care. A biased RIS is not a stand-alone fed antenna; it is a passive scatterer whose measured response depends on feed position, polarization, fixture scattering, reference state, and the applied control voltages. The control state must therefore be treated as part of the measurement definition, not as a secondary implementation detail.

The research gap addressed here is therefore not the invention of a new unit cell, but the reproducible measurement of RIS beam steering as a full-sphere scattering problem. A RIS measurement methodology should (i) preserve a controlled RF environment, (ii) scan the three-dimensional response rather than a single plane, (iii) program the RIS state without entering the chamber, (iv) record a reference state under identical scattering conditions, and (v) report relative gain and pointing metrics that can be recomputed from the measured angular data. This paper presents such a methodology and validates it experimentally on a varactor-based RIS prototype.

The main contributions are as follows. First, an automated anechoic-chamber methodology is introduced for full-sphere RIS measurements using a low-scattering fixture and synchronized RF acquisition and beam-steering programming. Second, a beam-steering synthesis procedure is defined for a grouped \(3\times3\) RIS control lattice using a pre-measured phase-voltage curve of a \SI{5}{GHz} varactor metasurface. Third, a measurement-based interpretation of RIS beam-steering gain is provided: the reported values are local, reference-normalized power ratios, and are therefore reported together with the angular window and pointing error. Fourth, the methodology is validated by a proof-of-operation campaign at \SI{5.3}{GHz}, where the largest target-window gain is \SI{8.8}{dB} and the largest maximum-window gain is \SI{14.0}{dB} after normalization to the unbiased-RIS reference.

\section{Measurement Platform}
\label{sec:platform}
The measurement system is designed around the spherical antenna-measurement chamber at TU Wien. The chamber supports measurements from \SI{800}{MHz} to \SI{40}{GHz}; its external dimensions are \SI{5.4}{m}\(\times\)\SI{5.1}{m}\(\times\)\SI{5.1}{m}. The interior is lined with \SI{500}{mm} pyramidal absorbers on the walls, ceiling, and part of the floor, yielding a central quiet zone of approximately \SI{1.5}{m^3} with a residual field level of about \SI{-40}{dB}. A Vector Telecom VT70HA18+SK horn antenna is used as excitation antenna \cite{VectorTelecomVT70HA18SK} which is mounted approximately \SI{850}{mm} from the RIS aperture. The RIS-exciter assembly is mounted on the chamber azimuth stage, while the receive probe is swept by the elevation positioner. The azimuth stage covers \(\phi\in[0^\circ,360^\circ)\), and the elevation sweep covers approximately \(\theta\in[-160^\circ,160^\circ]\), leaving a protected sector of about \SI{40}{\degree} without direct samples. The acquisition software exports a complete spherical grid; the missing sector is set to zero before the NSI NF--FF processing chain is applied. The resulting far-field data should therefore be interpreted as the transformed response of the complete RIS-exciter-fixture configuration under the stated zero-fill assumption.

\subsection{Device Under Test}
\label{subsec:dut}
\begin{figure*}[!h]
    \centering

    \begin{minipage}[b]{0.3\textwidth}
        \centering
        \includegraphics[width=\linewidth]{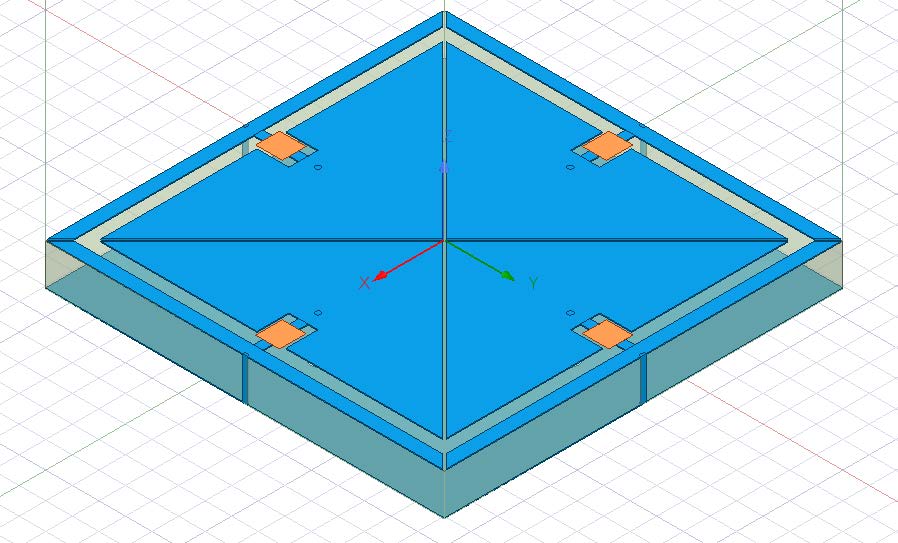}
        \vspace{1mm}
        \centerline{(a)}
    \end{minipage}
    \hspace{2cm}
    \begin{minipage}[b]{0.3\textwidth}
        \centering
        \includegraphics[width=\linewidth]{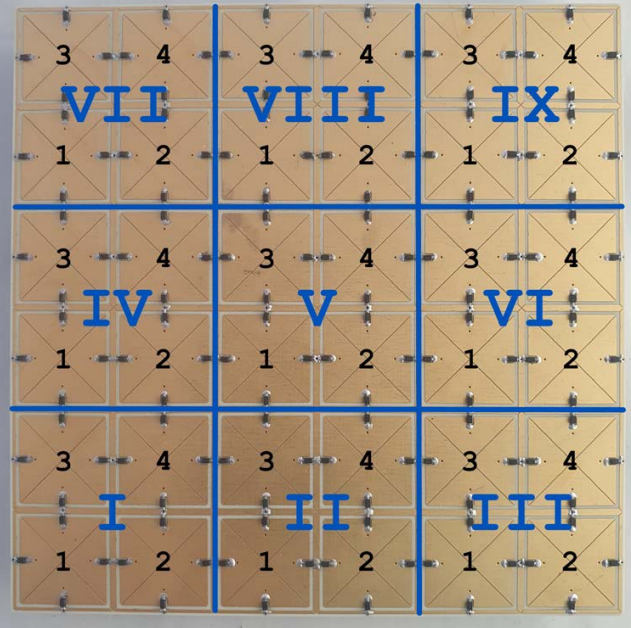}
        \vspace{1mm}
        \centerline{(b)}
    \end{minipage}

    \caption{Detailed view of the RIS hardware components:
    (a) RIS unit cell visualized in the simulation environment~\protect\cite{ritter2025};
    (b) complete RIS module with indication of subdivision grouping~\protect\cite{ritter2025}.}
    \label{fig:combined_ris_view}
\end{figure*}

The device under test is a modular varactor-based RIS designed for operation around \SI{5}{GHz}. It represents one instance of the broader electronically and materially tunable metasurface design space \cite{Molero2021MCOM,Basar2024VTM}. A single module contains a \(6\times6\) lattice of square unit cells on Rogers RO4003C substrate with \(\varepsilon_r=3.55\) and thickness \(t=3.2~\mathrm{mm}\). Each unit cell uses a square patch surrounded by a square; diagonal slots split the metallization, and four varactor diodes connect the patch and ring segments. The PCB is implemented as a four-layer stack-up so that the bias network and auxiliary connectors can be routed to the backside. This borderless layout permits later tiling of several modules without a mechanical gap at the active aperture. The physical module size is \(90~\mathrm{mm}\times90~\mathrm{mm}\), corresponding to approximately \(1.5\lambda\times1.5\lambda\) at \SI{5}{GHz}. The default control topology groups the \(36\) unit cells into nine \(2\times2\) subarrays. Each group is driven by one reverse-bias voltage, which reduces the control dimension to a \(3\times3\) programmable aperture. Prior characterization of this module showed an approximately \(326^{\circ}\) tunable reflection-phase range over the \SIrange{0}{20}{V} bias voltage interval. The RIS unit-cell design and the subarray configuration are depicted in Fig.~\ref{fig:combined_ris_view}  \cite{ritter2025}.

The grouped control topology is a deliberate compromise between wiring complexity and aperture programmability. Nine analog channels are sufficient to demonstrate anomalous reflection and to validate the measurement platform, but they also impose a finite spatial quantization of the phase profile. Consequently, the reported patterns should be interpreted as grouped-aperture beam-steering results rather than as the limiting performance of a fully addressable RIS.

\subsection{Low-Scattering Mechanical Fixture}
\label{subsec:mechanics}
The mechanical structure designed minimizes parasitic effects on both sides, near the DUT and the exicter-antenna. The carrier beam and DUT clamps are fabricated from ROHACELL 31 HF, while PTFE or other plastic fasteners replace metal hardware wherever possible. ROHACELL 31 HF is a closed-cell polymethacrylimide foam intended for antenna applications; representative material data give \(\varepsilon_r\approx1.04\) and \(\tan\delta\approx0.0016\) at \SI{5}{GHz} \cite{Evonik2018RohacellDielectric}. The carrier plate measures \SI{1250}{mm}\(\times\)\SI{625}{mm}\(\times\)\SI{55}{mm} and serves as the common mechanical reference for the exciter antenna, RIS clamps, and electronics plate. The RIS fixture provides repeatable edge indexing for the PCB and supports alternative apertures, including future multi-module configurations. The transmit horn, RIS, and controller electronics are carried by the same fixture so that the illumination geometry is preserved during the spherical scan.

\begin{figure*}[h!]
    \centering

    \begin{minipage}[b]{0.43\textwidth}
        \centering
        \includegraphics[width=\linewidth]{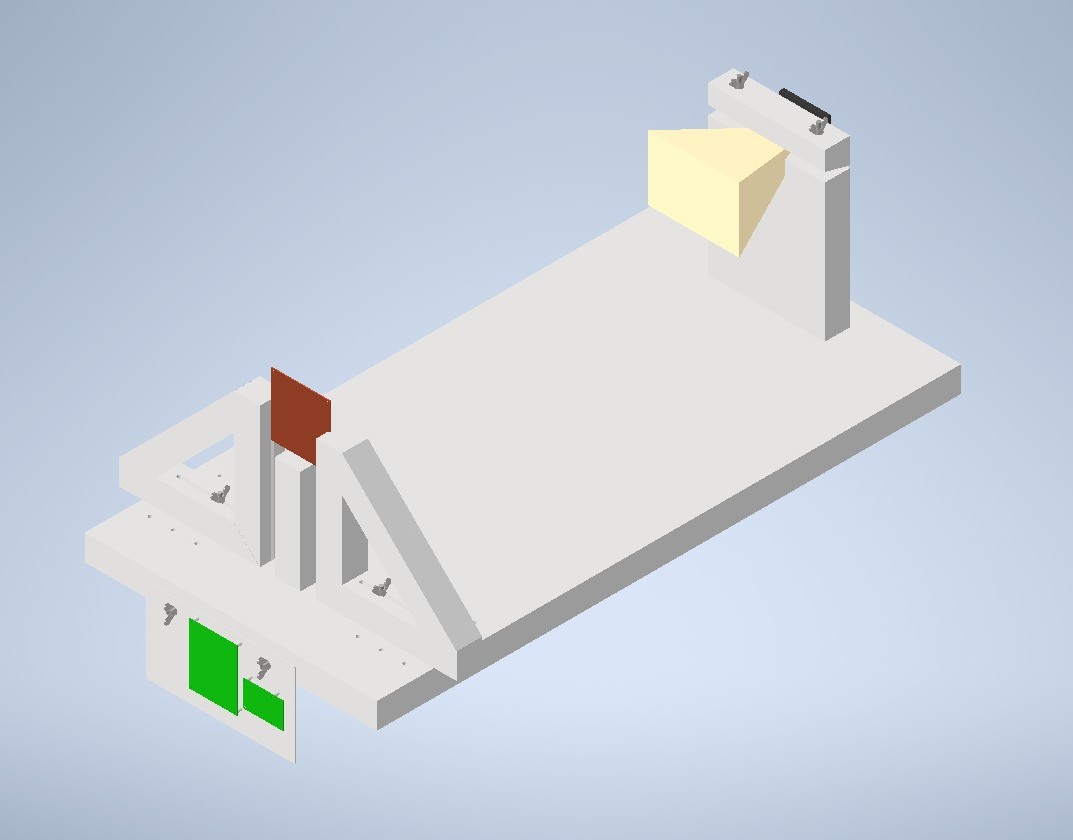}
        \vspace{1mm}
        \centerline{(a)}
    \end{minipage}
    \hfill
    \begin{minipage}[b]{0.43\textwidth}
        \centering
        \includegraphics[width=\linewidth]{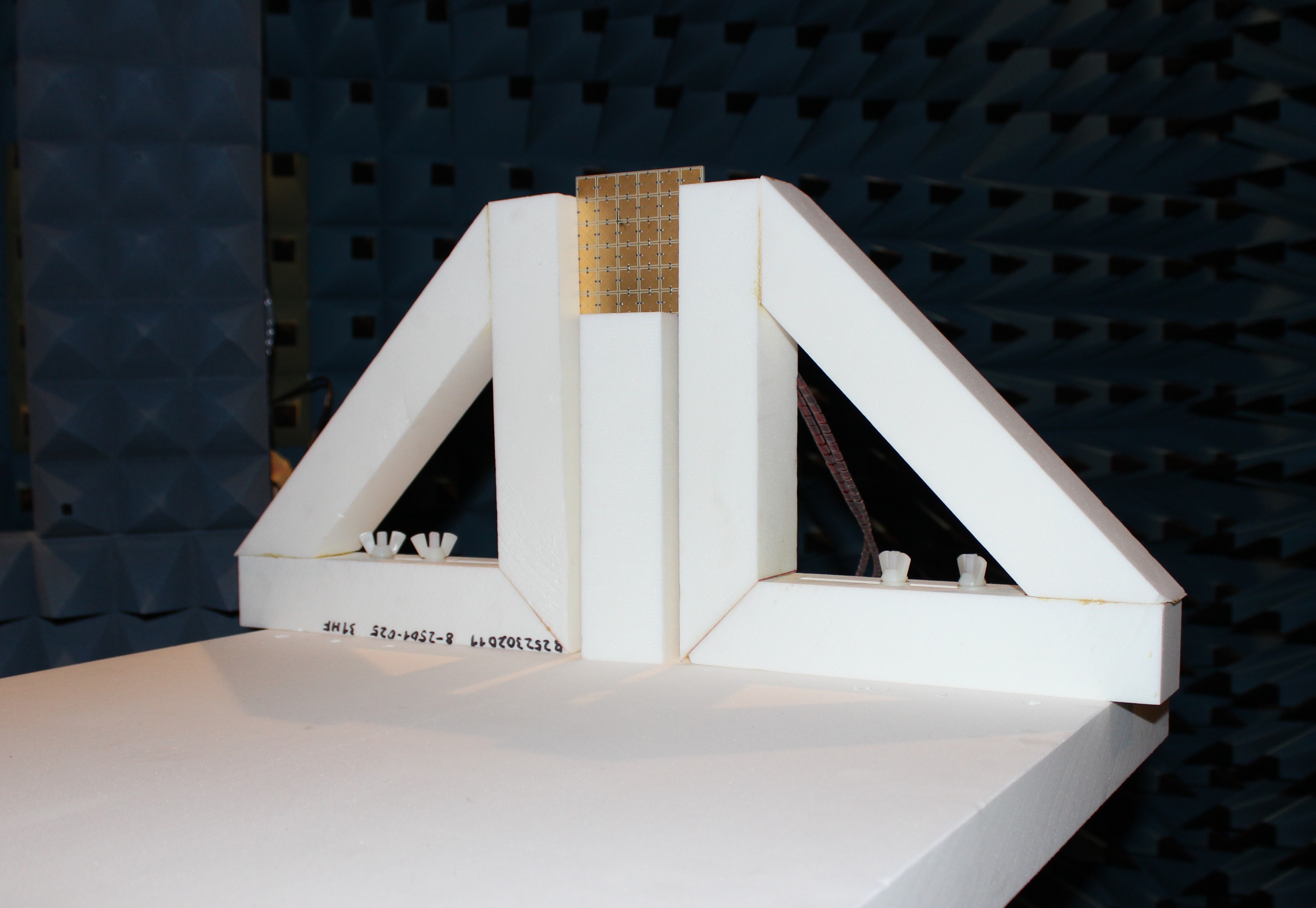}
        \vspace{1mm}
        \centerline{(b)}
    \end{minipage}

    \caption{RIS measurement fixture and modular assembly for anechoic chamber measurements:
    (a) modular RIS-antenna fixture showing the ROHACELL carrier plate, antenna fixture, RIS clamp, and electronics mounting plate. The RIS is highlighted in red, the DAC and Raspberry~Pi in green, and the exciter antenna in yellow;
    (b) RIS fixture with clamping elements on both sides of the RIS and the support column below.}
    \label{fig:ris_fixture_assembly}
\end{figure*}

The in-chamber electronics are mounted on a PMMA plate below the main beam of the transmit horn. This location shortens bias leads to the RIS while reducing exposure of the DAC board and Raspberry Pi controller to high incident fields. Reference measurements with and without an RF absorber sheet show that the electronics plate contributes measurable fine-structure to the background; for consistency, the beam-steering measurements are normalized to the reference state recorded with the same electronics installed and no absorber. The final setup is shown in Fig.~\ref{fig:ris_fixture_assembly}.

\subsection{Remote Bias Control and State Sequencing}
\label{subsec:bias_control}
\begin{figure*}[!h]
\centering
\includegraphics[width=0.9\textwidth]{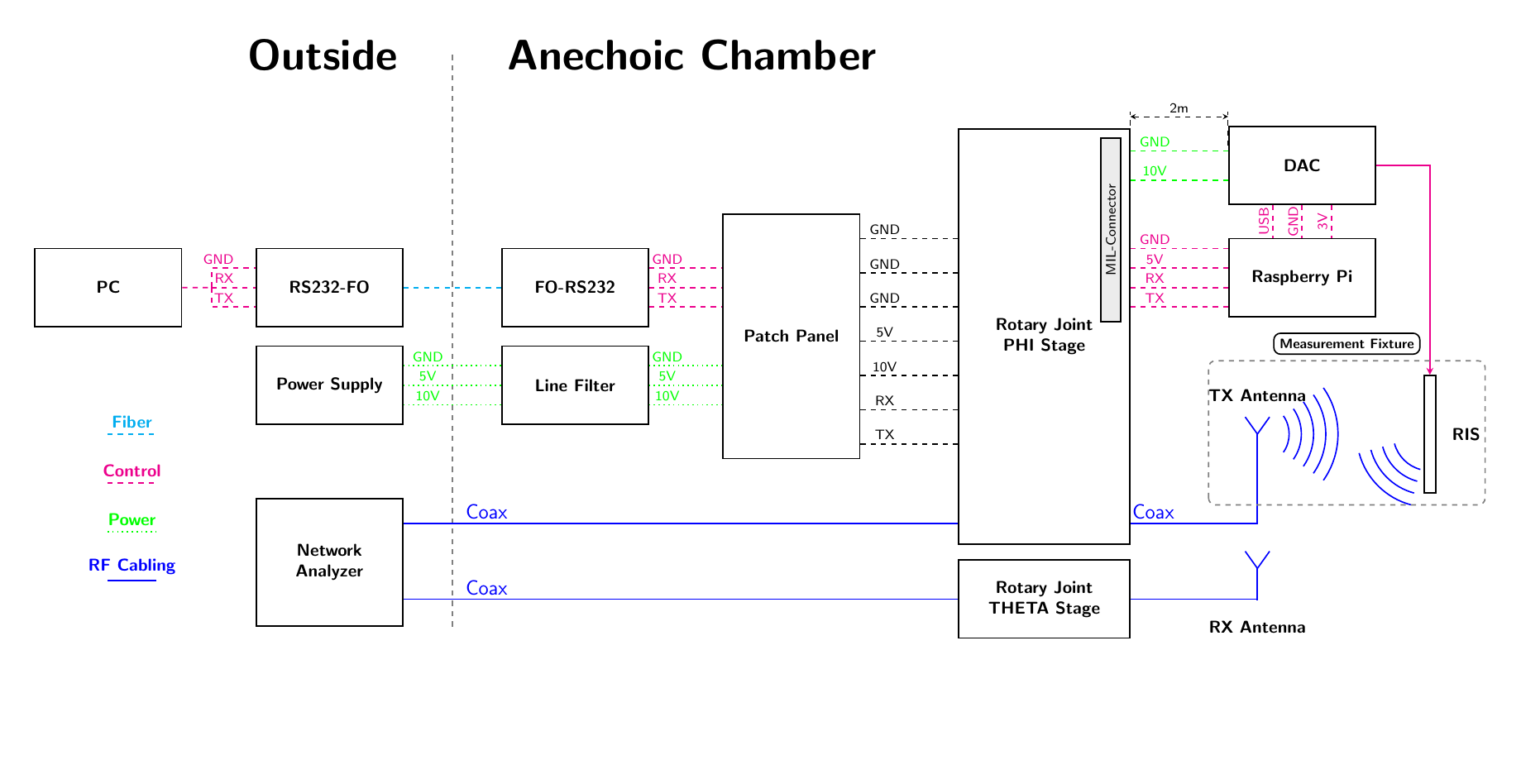}
\caption{Automated full-sphere RIS measurement platform. The essential feature is the combined workflow of mechanical scanning, complex \(S_{21}\) acquisition, and remote RIS bias programming without chamber access between states.}
\label{fig:setup}
\end{figure*}

Remote bias control is implemented by a Raspberry~Pi inside the chamber and a Texas Instruments DAC81416-08EVM digital-to-analog-converter board. The DAC provides \(16\) analog outputs. Nine channels are assigned to the nine RIS groups, and one channel (OUT15) is used as a \SI{3}{V} voltage-present flag monitored by a GPIO of the Raspberry~Pi. The DAC is connected to the Raspberry~Pi through micro-USB, which also provides the board-level \SI{5}{V} supply and serial communication. The Raspberry~Pi starts the DAC-control service automatically, receives configuration matrices over the chamber serial link, checks the matrix format, and returns error messages to the host PC if errors occur. A valid measurement state is therefore defined by the mechanical angles and the nine applied beam-steering voltages. An overview of the entire setup is given in Fig.~\ref{fig:setup}.

\section{Beam-Steering Methodology}
\label{sec:methodology}
This section formalizes the beam-state synthesis and the evaluation metrics used by the measurement methodology. The approach is intentionally simple: it uses a grouped planar-array phase law to command the RIS and then relies on full-sphere measurements to quantify the actual response. This separation is useful because deviations from the command reveal practical effects such as near-field illumination, mutual coupling, finite aperture, and phase-voltage interpolation error.

\subsection{Phase Gradient for Grouped RIS Control}
\label{subsec:phase_gradient}
The nine controllable RIS groups are modeled as a uniform \(3\times3\) planar array with group spacings \(d_y=d_z=30~\mathrm{mm}\). For operating wavelength \(\lambda\) and wavenumber \(k=2\pi/\lambda\), the inter-group phase increments required to steer the reflected beam toward the RIS-centered target direction \((\phi_t,\theta_t)\) are
\begin{align}
\Delta \phi_y &= -k d_y \cos(\theta_t)\sin(\phi_t), \\
\Delta \phi_z &= -k d_z \sin(\theta_t),
\label{eq:phase_increment}
\end{align}
where \(\phi_t\) is the azimuth angle and \(\theta_t\) is the elevation angle. For group indices \((m,n)\in\{-1,0,1\}^2\), the commanded phase is
\begin{equation}
\phi_{m,n}=\left(m\Delta\phi_y+n\Delta\phi_z+\phi_0\right) \bmod 2\pi ,
\label{eq:phase_profile}
\end{equation}
where \(\phi_0\) is an arbitrary common phase. In the measurements, \(\phi_0\) is selected implicitly by the phase-voltage interpolation so that all commands remain within the available tuning range.

At \SI{5}{GHz}, \(\lambda\approx60~\mathrm{mm}\), so the \SI{30}{mm} group spacing is approximately \(\lambda/2\). This spacing is appropriate for suppressing grating lobes in the principal steering region, although the small \(3\times3\) effective aperture necessarily yields broad beams. The purpose of the method is therefore not high angular resolution; it is a reproducible mapping from commanded angle to a measurable anomalous-reflection state.

\subsection{Phase-to-Voltage Mapping}
\label{subsec:phase_voltage}
\begin{figure}[!t]
\centering
    \includegraphics[width=0.75\linewidth]{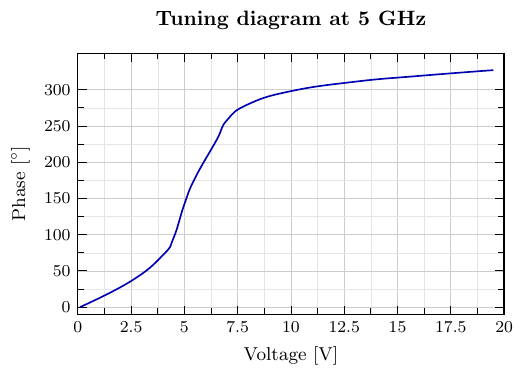}
\caption{Measured phase-voltage tuning characteristic used to interpolate the grouped RIS bias commands at \SI{5}{GHz}.}
\label{fig:tuning}
\end{figure}

The commanded phase matrix is converted to a voltage matrix using the measured phase-voltage tuning curve of the RIS module. Let \(g(V)\) denote the measured reflection phase at the operating frequency. The required bias voltage is obtained from the interpolated inverse map
\begin{equation}
V_{m,n}=g^{-1}\left(\phi_{m,n}\right), \quad (m,n)\in\{-1,0,1\}^2 .
\label{eq:voltage_map}
\end{equation}
The measured phase-voltage characteristic used for this inversion is shown in Fig.~\ref{fig:tuning}.
For example, the target \((\phi_t,\theta_t)=(50^{\circ},0^{\circ})\) at \SI{5}{GHz} gives \(\Delta\phi_y\approx-138.2^{\circ}\) and \(\Delta\phi_z=0^{\circ}\), resulting in the grouped bias matrix
\begin{equation}
\mathbf{V}_{50,0}=\begin{bmatrix}
0.10 & 6.52 & 4.37\\
0.10 & 6.52 & 4.37\\
0.10 & 6.52 & 4.37
\end{bmatrix}\,\mathrm{V} .
\label{eq:example_voltage}
\end{equation}
The repeated rows in \eqref{eq:example_voltage} create a one-dimensional phase gradient. This is expected to sharpen the response in the azimuthal steering dimension while leaving a broader response in the orthogonal dimension.

\subsection{Reference Normalization and Beam-Steering Gain}
\label{subsec:gain_metric}
Each beam-steering state is compared with a reference scan obtained with the RIS installed, all groups unbiased, and the same in-chamber electronics present. This reference choice intentionally retains the passive specular reflection of the unbiased RIS and the residual scattering of the overall chamber environment, so that normalization isolates changes caused by the programmed anomalous-reflection state. Since the evaluation uses ratios between a programmed state and a reference state measured with the same RF chain and processing workflow, absolute power calibration is not required in this framework; the reference measurement defines the normalization baseline.

Let \(S_{21}^{\mathrm{beam}}(\phi_i,\theta_i)\) and \(S_{21}^{\mathrm{ref}}(\phi_i,\theta_i)\) denote the measured complex transmission coefficients at angular sample \((\phi_i,\theta_i)\). For an evaluation region \(A\), the local mean powers are defined as
\begin{align}
\bar{P}_{\mathrm{beam}}(A)&=\frac{1}{N_A}\sum_{(\phi_i,\theta_i)\in A}\left|S_{21}^{\mathrm{beam}}(\phi_i,\theta_i)\right|^2,\\
\bar{P}_{\mathrm{ref}}(A)&=\frac{1}{N_A}\sum_{(\phi_i,\theta_i)\in A}\left|S_{21}^{\mathrm{ref}}(\phi_i,\theta_i)\right|^2,
\label{eq:local_power}
\end{align}
where \(N_A\) is the number of samples in the region. The beam-steering gain is
\begin{equation}
G_{\mathrm{BS}}(A)=10\log_{10}\left(\frac{\bar{P}_{\mathrm{beam}}(A)}{\bar{P}_{\mathrm{ref}}(A)}\right)\,\mathrm{dB}.
\label{eq:gain}
\end{equation}
Two choices of \(A\) are used. The first is a local \(\pm5^\circ\) angular window centered at the target direction, which quantifies the performance where the beam was intended to appear. The second is a local window centered at the maximum of a smoothed normalized map, which quantifies the strongest  anomalous reflection achieved even when the beam is displaced from the command. The angular displacement is reported as the spherical separation between the target and the measured maximum.

\section{Experimental Results}
\label{sec:results}
The proof-of-operation campaign first establishes the fixture and chamber background and then evaluates beam-steering patterns over frequency and direction. Unless stated otherwise, results are shown after normalization to the unbiased-RIS reference described in Section~\ref{subsec:gain_metric}. Therefore, positive values indicate that the programmed RIS state increases the received power relative to the reference measurement in that angular region.

\subsection{Reference Measurements}
\label{subsec:reference}
Four baseline configurations were measured at \SI{5}{GHz}: transmit horn only, horn with in-chamber electronics installed, complete fixture with the unbiased RIS, and complete fixture with an additional sheet absorber near the electronics. The horn-only case shows the expected main radiation region and a measurable backlobe. Installing the controller electronics adds a fine interference structure. Adding the unbiased RIS produces a stronger local response near the specular sector, while the absorber sheet smooths part of the background caused by electronics scattering. The visualized horn-only reference measurement and the reference measurement of the complete setup, including the RIS, are shown in Fig.~\ref{fig:reference}. The remaining two reference measurements are provided in the appendix.

The absorber-assisted reference is not used for the beam-steering normalization because the actual beam measurements were performed without that absorber. Using it would mix two electromagnetic environments and would artificially alter the reported gain. The selected reference is therefore the complete fixture with the unbiased RIS and no absorber. This is conservative because it includes both passive RIS reflection and residual electronics scattering in the denominator of \eqref{eq:gain}.

\begin{figure*}[!t]
    \centering

    \begin{minipage}[b]{0.43\textwidth}
        \centering
        \includegraphics[width=\linewidth]{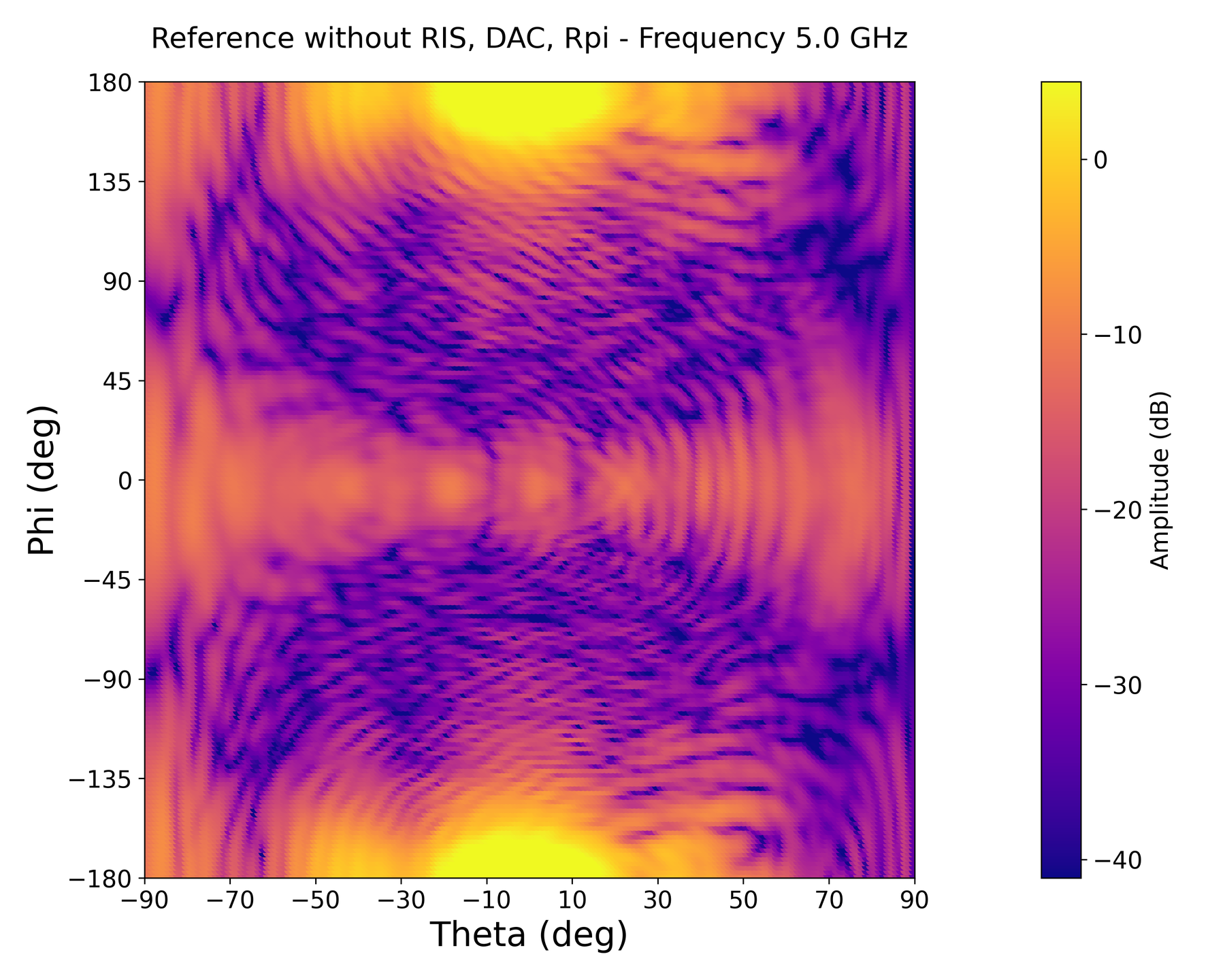}
        \vspace{1mm}
        \centerline{(a)}
        \label{fig:ref-a}
    \end{minipage}
    \hfill
    \begin{minipage}[b]{0.43\textwidth}
        \centering
        \includegraphics[width=\linewidth]{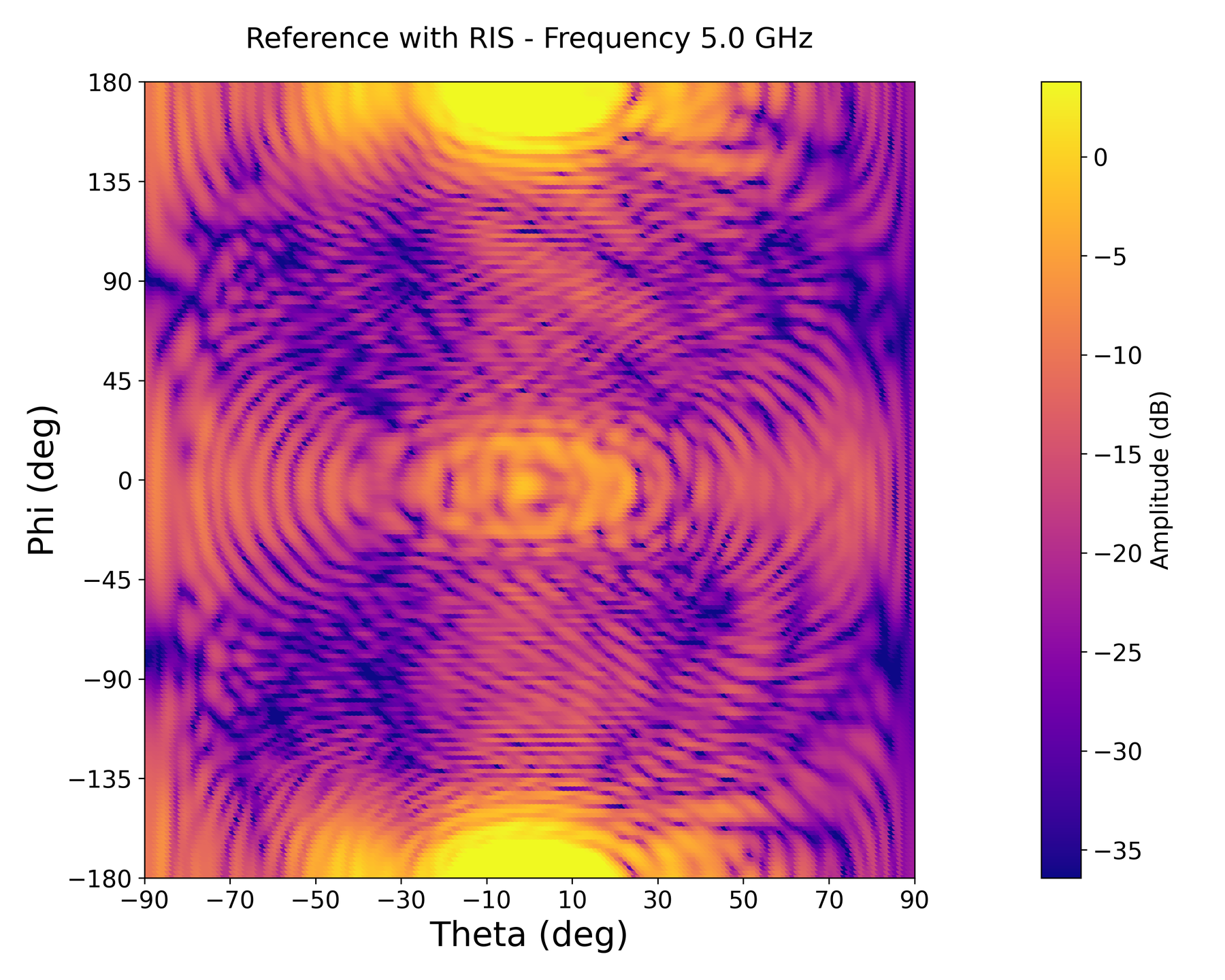}
        \vspace{1mm}
        \centerline{(b)}
        \label{fig:ref-b}
    \end{minipage}

    \caption{Reference measurements under four configurations:
    (a) reference without RIS, DAC, and Raspberry~Pi;
    (b) reference with RIS, with all groups at $V_{\mathrm{bias}}=0\,\mathrm{V}$.}
    \label{fig:reference}
\end{figure*}

\subsection{Frequency Dependence of a One-Dimensional Beam}
\label{subsec:freq}
\begin{figure*}[!t]
    \centering

    \begin{minipage}[b]{0.42\textwidth}
        \centering
        \includegraphics[width=\linewidth]{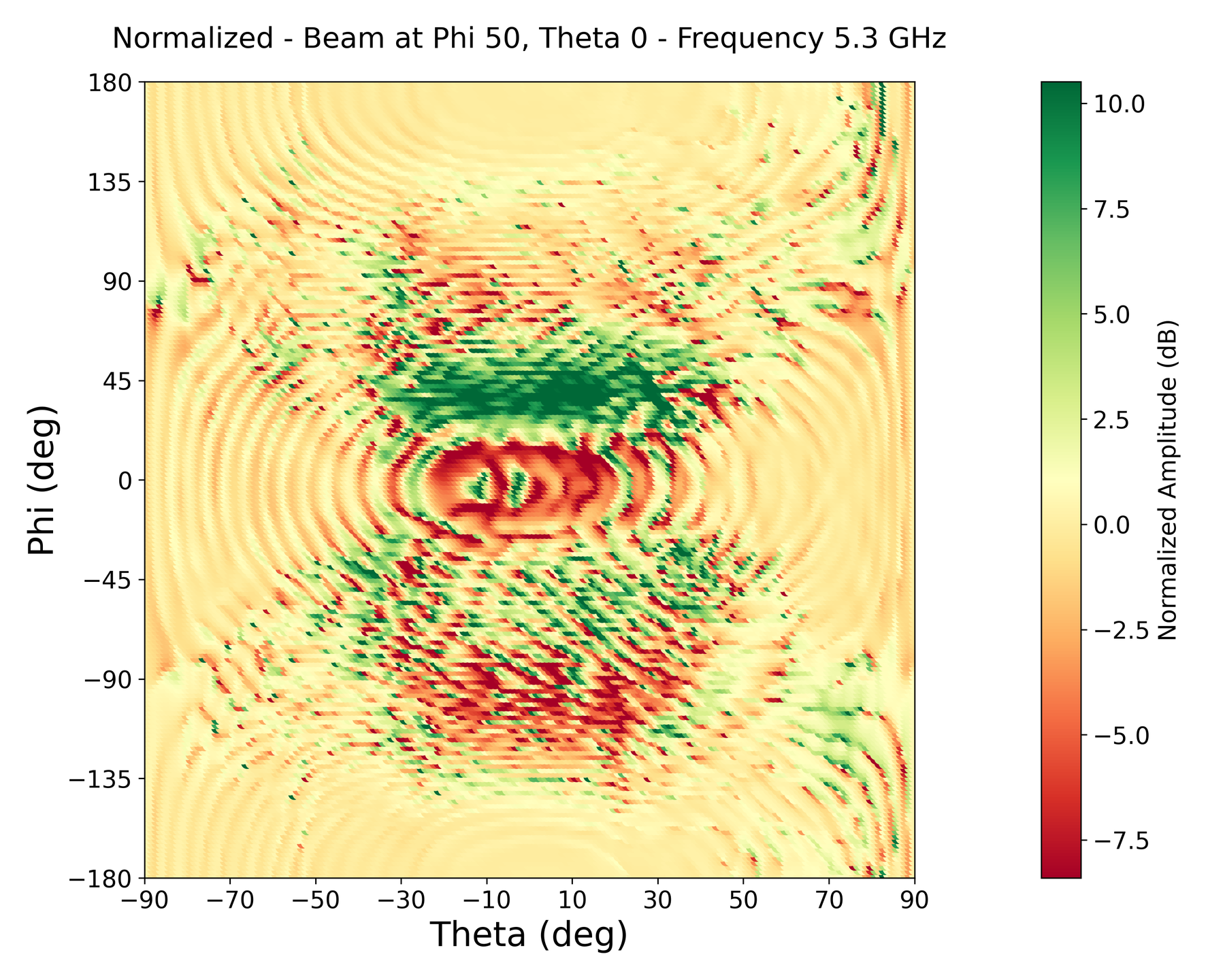}
        \vspace{1mm}
        \centerline{(a)}
        \label{fig:1Dim_5G3}
    \end{minipage}
    \hfill
    \begin{minipage}[b]{0.42\textwidth}
        \centering
        \includegraphics[width=\linewidth]{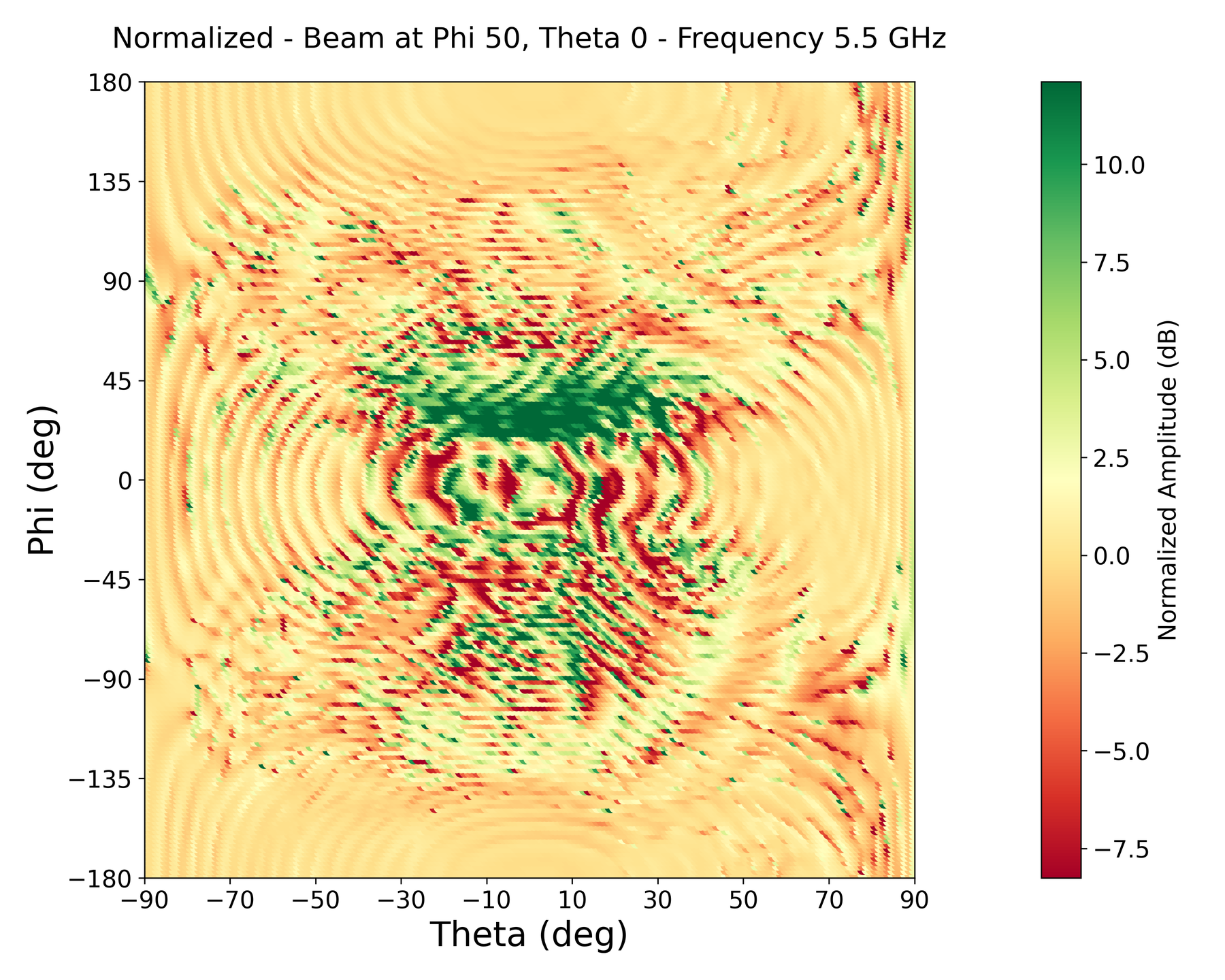}
        \vspace{1mm}
        \centerline{(b)}
    \end{minipage}

    \caption{Normalized radiation maps for the beam at $\phi=50^\circ$ and $\theta=0^\circ$ at representative frequencies:
    (a) 5.3~GHz;
    (b) 5.5~GHz. The intermediate frequency points are moved to the appendix.}
    \label{fig:data_1}
\end{figure*}

The beam-steering target \((\phi_t,\theta_t)=(50^{\circ},0^{\circ})\) was measured from \SIrange{5.0}{5.5}{GHz}. A clear anomalous-reflection region is visible over \SIrange{5.0}{5.4}{GHz}, with the cleanest main-lobe definition around \SI{5.3}{GHz}. Above \SI{5.4}{GHz}, the normalized maps exhibit stronger ripple and pattern distortion. This trend is consistent with using a phase-voltage tuning curve optimized near \SI{5}{GHz} and with the  change in electrical-size of the grouped aperture over the band. Representative maps at \SI{5.3}{GHz} and \SI{5.5}{GHz} for the one-dimensinal beam are shown in Fig.~\ref{fig:data_1}, while the remaining frequency points are moved to the appendix.

The pattern is broad in elevation for the one-dimensional command because the voltage matrix in \eqref{eq:example_voltage} repeats the same row three times. Physically, the aperture behaves like three replicated strip excitations: it provides a stronger phase gradient in the azimuthal dimension than in elevation. This behavior is not a failure of the measurement; rather, it is a useful demonstration that the full-sphere scan captures the anisotropic beamwidth produced by the grouped control law.

\subsection{Two-Dimensional Beam Steering and Directional Accuracy}
\label{subsec:directional}
\begin{figure*}[t]
    \centering
    \begin{minipage}{0.43\textwidth}
        \centering
        \includegraphics[width=\linewidth]{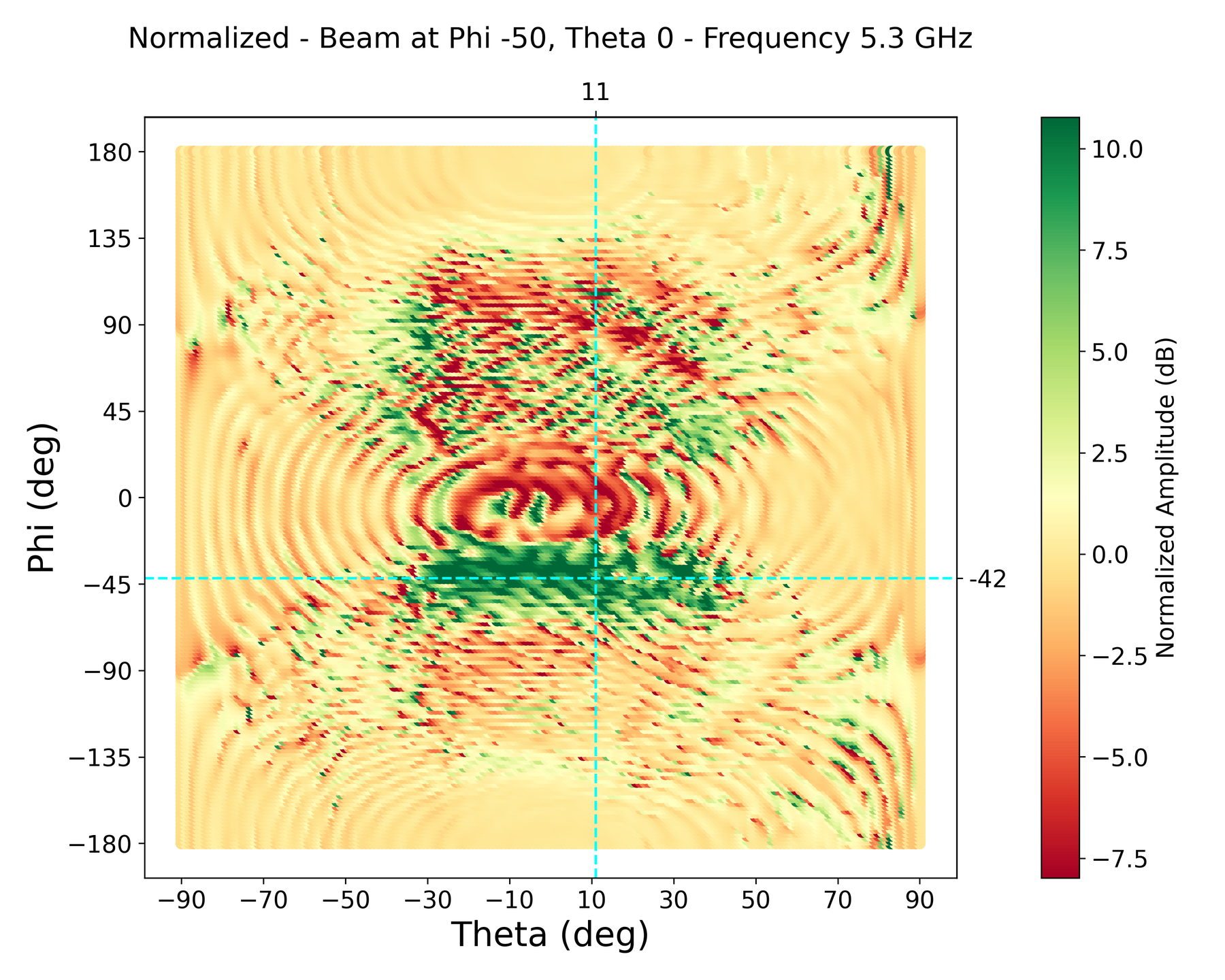}
        \centerline{\footnotesize (a) $\phi_\mathrm{t}=-50^\circ,\;\theta_\mathrm{t}=0^\circ$}
    \end{minipage}
    \hfill
    \begin{minipage}{0.43\textwidth}
        \centering
        \includegraphics[width=\linewidth]{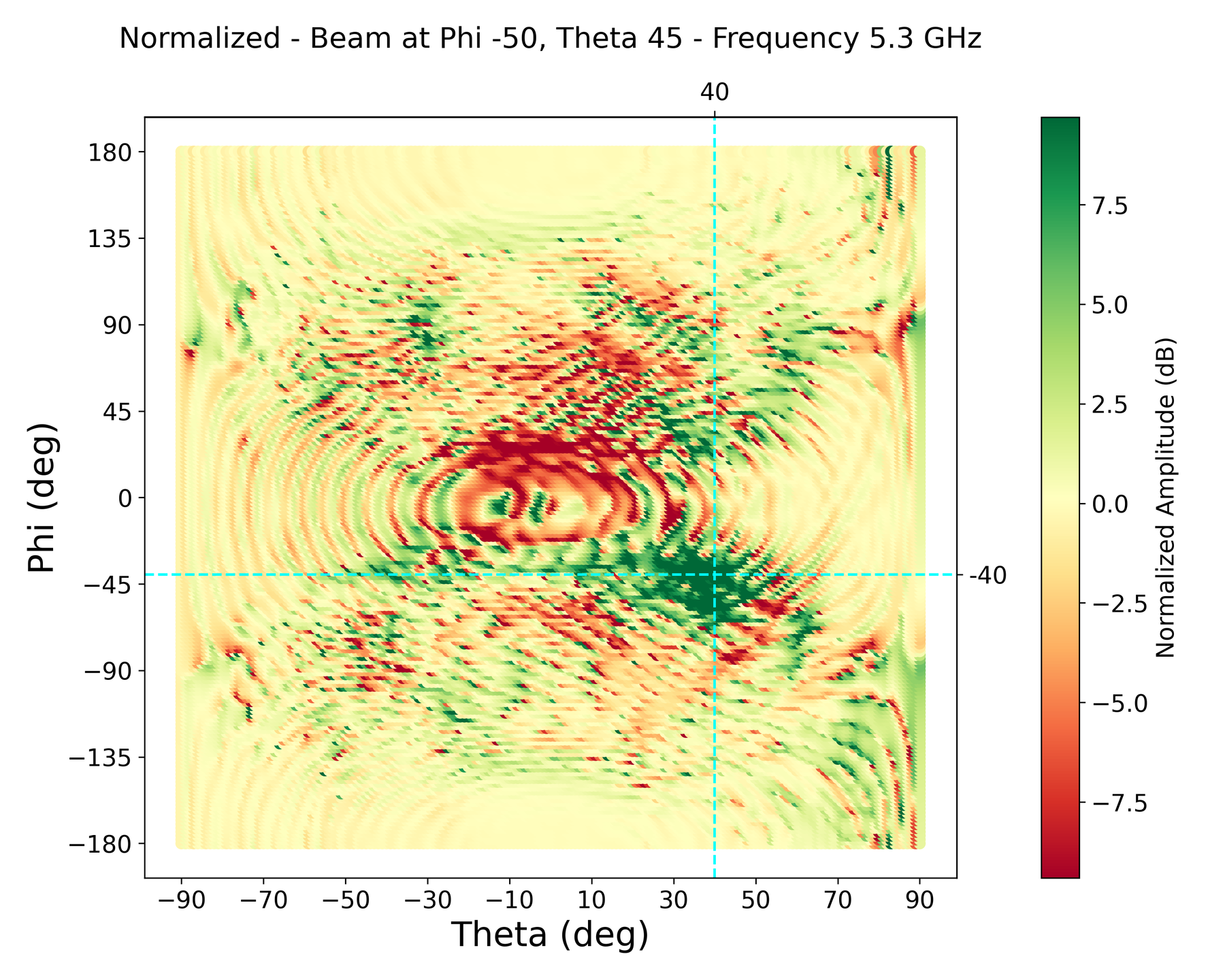}
        \centerline{\footnotesize (b) $\phi_\mathrm{t}=-50^\circ,\;\theta_\mathrm{t}=45^\circ$}
    \end{minipage}

    \vspace{1em}

    \begin{minipage}{0.43\textwidth}
        \centering
        \includegraphics[width=\linewidth]{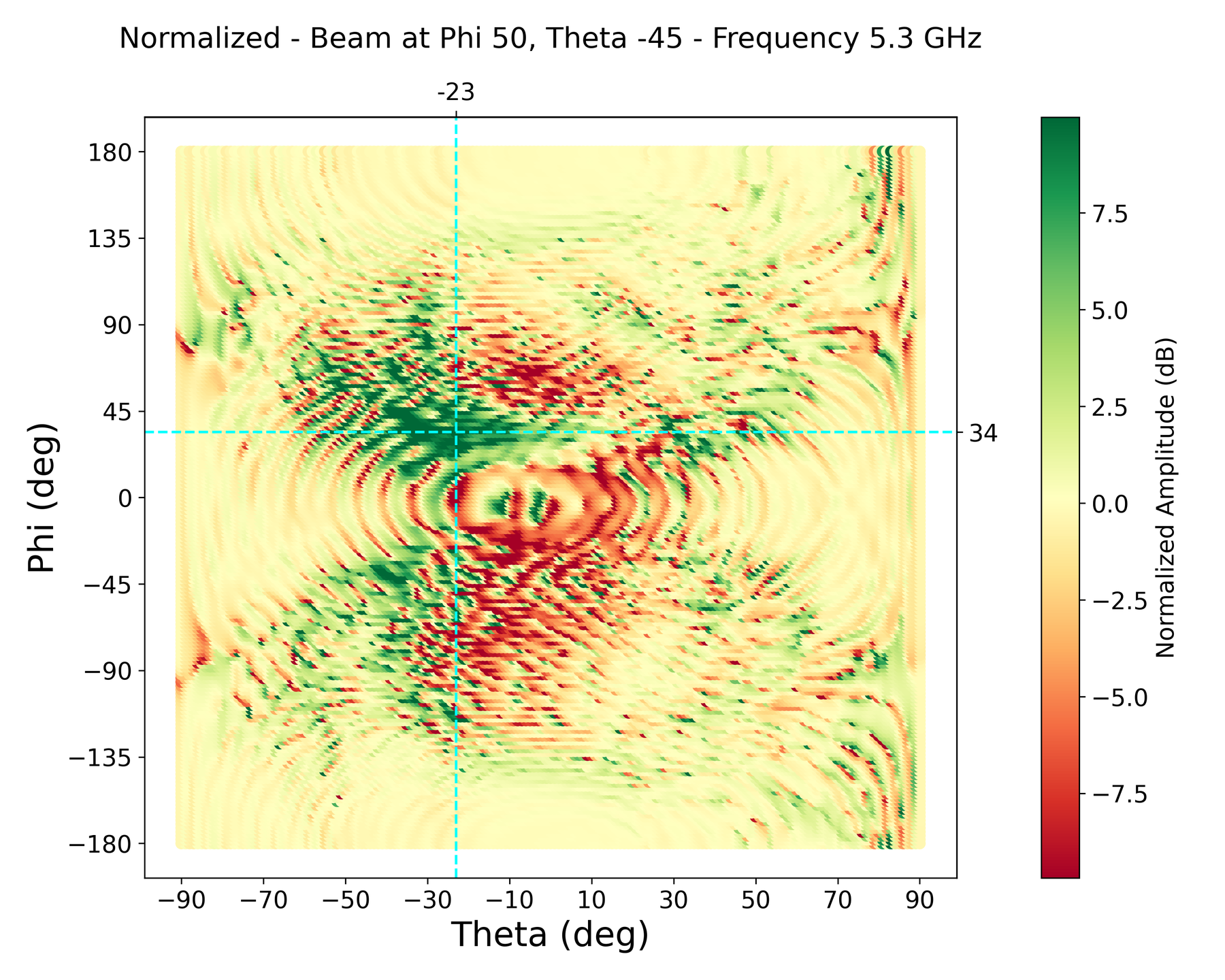}
        \centerline{\footnotesize (c) $\phi_\mathrm{t}=50^\circ,\;\theta_\mathrm{t}=-45^\circ$}
    \end{minipage}
    \hfill
    \begin{minipage}{0.43\textwidth}
        \centering
        \includegraphics[width=\linewidth]{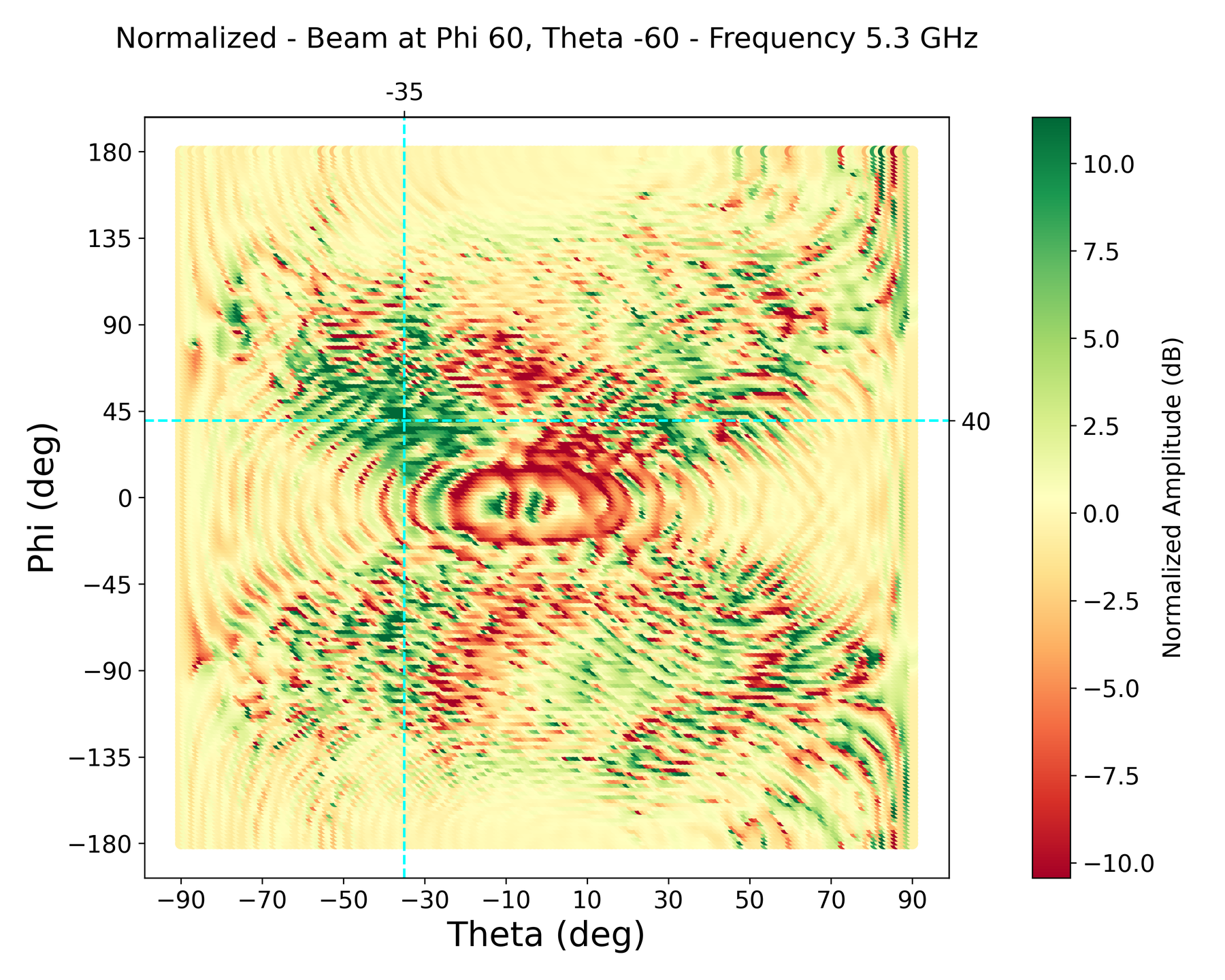}
        \centerline{\footnotesize (d) $\phi_\mathrm{t}=60^\circ,\;\theta_\mathrm{t}=-60^\circ$}
    \end{minipage}
    \caption{Normalized radiation maps of different RIS beam configurations at $5.3\,\mathrm{GHz}$, with highlighted maximum coordinates.}

\label{fig:maxima}
\end{figure*}

Measurements at \SI{5.3}{GHz} demonstrate that the methodology can program, measure, and localize both one-dimensional and two-dimensional steering states. The one-dimensional commands \((50^{\circ},0^{\circ})\) and \((-50^{\circ},0^{\circ})\) produce symmetric responses with target-window gains of \SI{8.0}{dB} and \SI{7.0}{dB}, respectively. The two-dimensional command \((-50^{\circ},45^{\circ})\) gives the strongest result, with a target-window gain of \SI{8.8}{dB} and a measured maximum-window gain of \SI{14.0}{dB}. The commands \((50^{\circ},-45^{\circ})\) and \((60^{\circ},-60^{\circ})\) still produce positive target-window gain but show larger pointing offsets. Fig.~\ref{fig:maxima} shows the normalized radiation maps for the one-dimensional beam \((-50^{\circ},0^{\circ})\) and the two-dimensional beams at \SI{5.3}{GHz}.

Table~\ref{tab:beam_summary} consolidates the two gain definitions and the measured pointing offsets. Averaged over the five measured beam states, the target-window gain is \SI{5.6}{dB}, whereas the maximum-window gain is \SI{11.3}{dB}. The difference between these metrics is the central measurement result: the programmed RIS states create local power enhancement relative to the unbiased reference, but the achieved peak can be displaced from the analytical command. This is expected for a grouped aperture illuminated in the radiating near field and synthesized using a simplified plane-wave phase law.

\begin{table*}[!t]
\caption{Relative Beam-Steering Gain at \SI{5.3}{GHz} Using the Unbiased-RIS Reference}
\label{tab:beam_summary}
\centering
\begin{tabular}{
    S[table-format=-2.0]
    S[table-format=-2.0]
    c
    S[table-format=2.1]
    S[table-format=-2.0]
    S[table-format=-2.0]
    S[
        table-format=2.1,
        table-column-width=3.8em,
        table-number-alignment=right
    ]
    @{\,/\,}
    S[
        table-format=2.1,
        table-column-width=3.8em,
        table-number-alignment=left
    ]
}
\toprule
\multicolumn{2}{c}{Commanded target} & Steering & {Target-window} & \multicolumn{2}{c}{Measured maximum} & \multicolumn{2}{c}{Maximum-window} \\
\cmidrule(lr){1-2}\cmidrule(lr){5-6}\cmidrule(lr){7-8}
{\(\phi_t\)} & {\(\theta_t\)} & type & {gain (dB)} & {\(\phi_{\max}\)} & {\(\theta_{\max}\)} & \multicolumn{2}{c}{gain / error} \\
{(deg)} & {(deg)} & & {\(G_{\mathrm{BS}}(A_t)\)} & {(deg)} & {(deg)} & \multicolumn{2}{c}{(dB / deg)} \\
\midrule
 50 &   0 & 1-D & 8.0 &  42 &   9 & 12.1 & 12.0 \\
-50 &   0 & 1-D & 7.0 & -42 &  11 & 10.2 & 13.6 \\
-50 &  45 & 2-D & 8.8 & -40 &  40 & 14.0 &  8.9 \\
 50 & -45 & 2-D & 2.5 &  34 & -23 & 10.6 & 25.6 \\
 60 & -60 & 2-D & 1.8 &  40 & -35 &  9.4 & 28.2 \\
\midrule
\multicolumn{3}{r}{Mean over five states} & 5.6 & \multicolumn{2}{c}{} & 11.3 & 17.7 \\
\bottomrule
\end{tabular}
\end{table*}

\subsection{Comparison With Representative Measurement Campaigns}
\label{subsec:comparison}
Table~\ref{tab:literature_comparison} positions the proposed methodology relative to representative RIS measurement studies. The comparison is qualitative because reported gains, path loss, and channel coefficients are normalized differently across studies. The main distinction is that the present platform combines full-sphere pattern acquisition with synchronized bias programming of the RIS state. This makes it suitable for extracting beam pointing, sidelobes, and off-plane scattering features that a single angular cut may miss.

\begin{table}[!t]
\caption{Positioning of the Proposed RIS Measurement Methodology}
\label{tab:literature_comparison}
\centering
\scriptsize
\begin{tabular}{L{0.30\linewidth} L{0.62\linewidth}}
\toprule
Study & Methodological distinction \\
\midrule
Tang \emph{et al.} \cite{Tang2021TWC} & Fixed bistatic path-loss validation at \SI{4.25}{GHz} and \SI{10.5}{GHz}; no full-sphere beam map for each programmed state. \\
Tewes \emph{et al.} \cite{Tewes2023VTC} & Reproducible \SIrange{5}{6}{GHz} channel dataset with angular and scenario variation; not a synchronized spherical scan of every RIS beam state. \\
Trichopoulos \emph{et al.} \cite{Trichopoulos2022OJCOMS} & Chamber and outdoor prototype evaluation at \SI{5.8}{GHz}; reports link improvement rather than a metrological full-sphere scattering response. \\
This work & Near-complete spherical acquisition, complex \(S_{21}\), synchronized 9-channel analog bias, relative target-window gain, maximum-window gain, and pointing error extracted from one dataset. \\
\bottomrule
\end{tabular}
\end{table}

% \clearpage

\section{Discussion}
\label{sec:discussion}
The measured results confirm anomalous reflection, but their main value is methodological. The platform does not merely record a received power level in one selected link geometry. It records a state-dependent spherical scattering response and therefore makes beam displacement, angular spread, off-plane lobes, and fixture-induced artifacts visible in the same coordinate system. This is the measurement capability that is still weakly established in the RIS literature.

The interpretation of ``gain'' requires strict wording. The investigated RIS is passive in the RF path; it does not generate an active power gain. The values in Table~\ref{tab:beam_summary} are therefore not amplifier gain, not realized antenna gain of the RIS, and not a geometry-independent link-budget parameter. They are local received-power ratios between two complete scattering states measured with the same antennas, fixture, chamber, cables, and processing chain. A positive value of \(G_{\mathrm{BS}}(A)\) means that the programmed bias state delivers more power into the angular region \(A\) than the unbiased reference state. This does not imply that the total scattered power has increased.  This can also be seen directly in the presented plots: apart from the intended areas of increased power shown in green, there are always distinct red areas indicating a reduction in available power. For a passive RIS, local gain is primarily a redistribution metric.

This distinction also explains why two gain windows are reported. The target-window gain answers the communication-oriented question: how much additional power appears where the RIS was commanded to steer the beam? The maximum-window gain answers a different diagnostic question: where did the strongest programmed scattering response actually occur, and how large was it relative to the reference? Reporting only the maximum-window gain would overstate practical steering quality when the peak is displaced from the target. Reporting only the target-window gain would hide the fact that the surface did form a stronger beam, but not exactly at the commanded direction. The paired metric and the angular error are therefore part of the methodology, not just post-processing.

The reference state is equally important. The unbiased-RIS reference used here includes the passive specular reflection of the RIS, the fixture, the electronics plate, and the same cable routing. This is a conservative reference for beam steering because the denominator already contains the passive scatterer. Other references, such as a metal plate, an absorber-loaded fixture, or a no-RIS configuration, would lead to different numerical values. For this reason, the gain values in Table~\ref{tab:beam_summary} should not be compared directly with gains reported in unrelated RIS campaigns unless the reference, angular window, calibration plane, and processing chain are equivalent.

A full-sphere dataset also allows for an additional consistency check that is unavailable in single-cut measurements. For angular weights \(w_i\) representing the sampled solid angle, one may define an integrated redistribution ratio
\begin{equation}
R_{\Omega}=10\log_{10}\left(\frac{\sum_i w_i\left|S_{21}^{\mathrm{beam}}(\phi_i,\theta_i)\right|^2}{\sum_i w_i\left|S_{21}^{\mathrm{ref}}(\phi_i,\theta_i)\right|^2}\right)\,\mathrm{dB}.
\label{eq:integrated_ratio}
\end{equation}
This quantity is not used as an efficiency estimate in the present paper because the setup is optimized for reproducible pattern comparison, not for absolute scattered-power calibration, and because the spherical scan contains a mechanically unavailable sector. It is nevertheless a useful future metric: a strong local gain together with a bounded integrated ratio would document that the RIS redirects energy rather than amplifying it.

The observed pointing offsets follow from the assumptions in the steering law. Equations~\eqref{eq:phase_increment}--\eqref{eq:phase_profile} assume a planar grouped aperture, plane-wave incidence, and phase-only reflection. The actual experiment includes near-field illumination from the transmit horn, phase-dependent amplitude variation of the varactor-loaded cells, mutual coupling between neighboring unit cells, finite-aperture diffraction, and scattering from support structures. Both, practical phase-shift models and electromagnetic-compliant mutual-coupling models predict this type of deviation from the ideal phase-only beamforming \cite{Abeywickrama2020TCOMM,Gradoni2021LWC}.

Several improvements follow directly from the measurement evidence. Per-cell control would reduce quantization error and allow phase profiles closer to the continuous solution. A near-field-aware synthesis model should replace \eqref{eq:phase_increment} when the feed is close to the aperture \cite{Jiang2023TWC,Mu2024VTM}. Calibration of the complex reflection coefficient, including phase-dependent amplitude, would improve voltage selection. The zero-filled missing sector in the spherical scan should also be treated as a limitation when interpreting sidelobes close to the unavailable angular region. Finally, local absorber treatment or further fixture improvement can reduce support-column and electronics scattering without altering the intended RIS aperture response.

\section{Conclusion}
\label{sec:conclusion}
This paper presented an automated full-sphere measurement methodology for a varactor-based sub-6 GHz RIS in an anechoic chamber. The contribution is the coupled workflow: low-scattering mechanical support, remote analog bias control,  complex \(S_{21}\) acquisition, near-field-to-far-field processing, unbiased-state normalization, and paired extraction of target-window gain, maximum-window gain, and pointing error. A grouped \(3\times3\) phase-gradient synthesis was applied to a \SI{5}{GHz} varactor RIS as a proof of operation. At \SI{5.3}{GHz}, the measured beam states achieved up to \SI{8.8}{dB} relative gain in the commanded target window and up to \SI{14.0}{dB} in the measured maximum window. These values quantify local power redistribution relative to the unbiased RIS; they should not be read as active gain. The observed pointing offsets motivate near-field-aware synthesis, per-cell control, and improved fixture scattering suppression. The platform provides a reproducible basis for comparing RIS prototypes beyond single-plane or fixed-bistatic measurements.

% Keep the appendix figures before the back matter.
\FloatBarrier
\section*{Acknowledgment}
This work has been funded by the Christian Doppler Laboratory for Digital Twin Assisted AI for Sustainable Radio Access Networks. The financial support by the Austrian Federal Ministry for Labour and Economy, the National Foundation for Research, Technology and Development, and the Christian Doppler Research Association is gratefully acknowledged.

\bibliographystyle{IEEEtran}
\bibliography{references}

% IEEE Access requires a short biography for every author.
\begin{IEEEbiography}[{\includegraphics[width=1in,height=1.25in,clip,keepaspectratio]{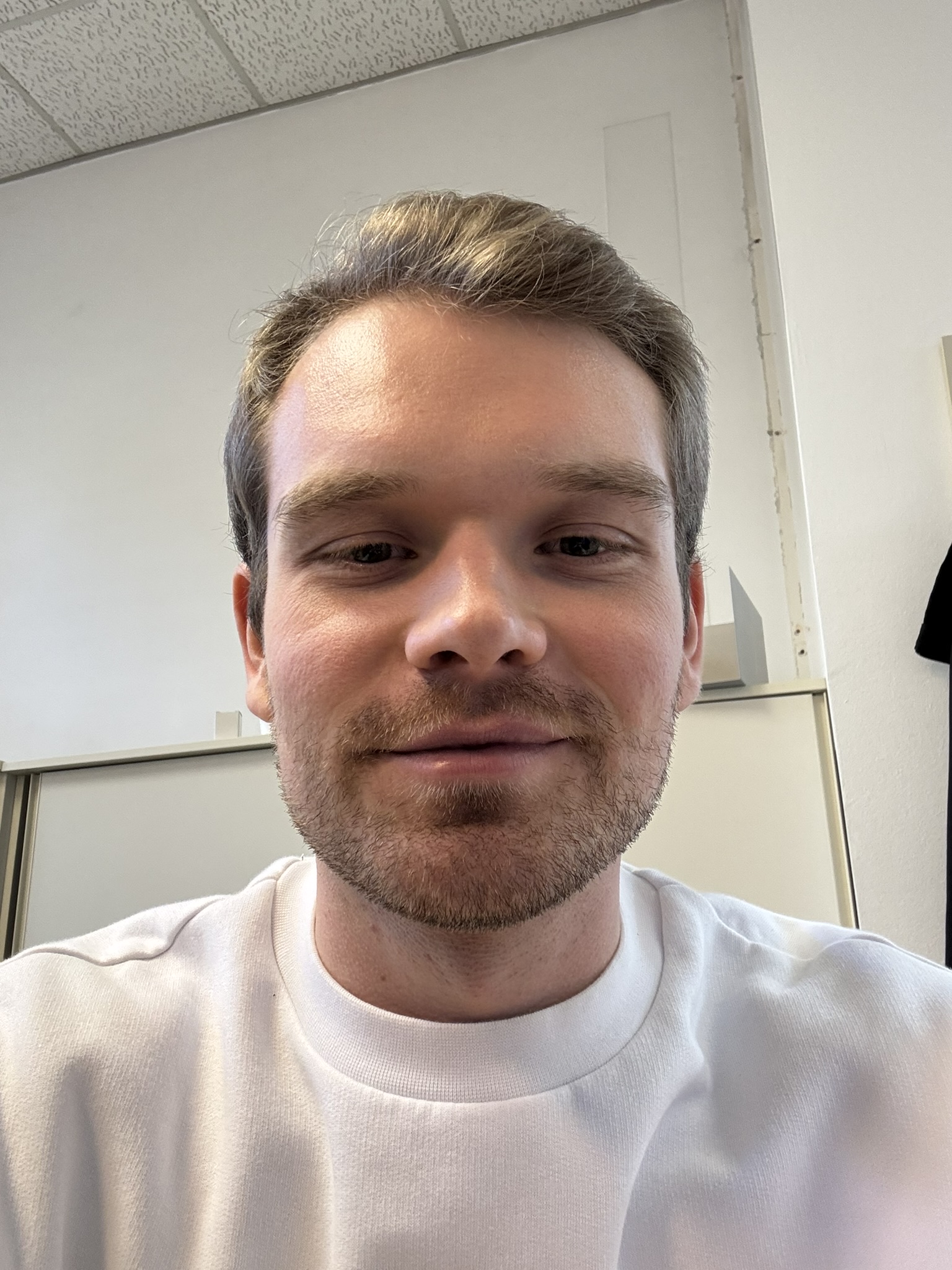}}]{Tobias Kancz}, Tobias Kancz received the B.Sc. degree from TU Wien, Vienna, Austria, where his bachelor’s thesis focused on the automated measurement and characterization of reconfigurable intelligent surfaces. He is currently pursuing the master’s degree in telecommunications at TU Wien and works with the Christian Doppler Laboratory for Digital Twin Assisted AI for Sustainable Radio Access Networks. His research interests include reconfigurable intelligent surfaces, automated radio-frequency measurement systems, antenna characterization, and experimental methods for wireless communication systems.
\end{IEEEbiography}
% Add Tobias Kancz's verified biography and portrait here before journal submission.
\begin{IEEEbiography}[{\includegraphics[width=1in,height=1.25in,clip,keepaspectratio]{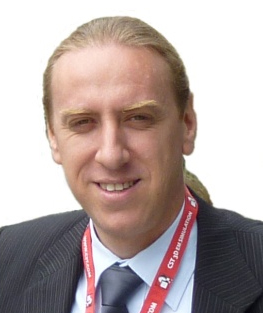}}]{Robert Langwieser}, Dr. techn. studied telecommunications at the Vienna University of Technology (TU Wien) and earned his Master degree 2004 and his Doctor degree in 2009, both with distinction.
From February 2009 until December 2012 he was post-doc project assistant at the same institute and module leader at two Christian Doppler Laboratorys.
Since January 2013 Dr. Langwieser is employed as Senior Scientist at the same institute and is responsible for the anechoic chamber for antenna measurements up to 40\,GHz.
His research and technical focus are on wireless systems, including antennas, microwave circuit and system design and simulation, and microwave measurement techniques.
\end{IEEEbiography}

\begin{IEEEbiography}[{\includegraphics[width=1in,height=1.25in,clip,keepaspectratio]{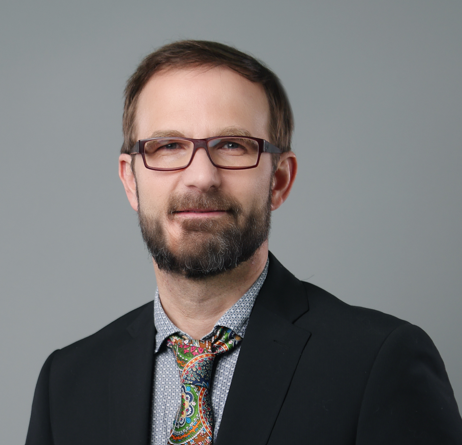}}]{Philipp Svoboda}
received the
Dr.Ing. degree in electrical engineering from Technische Universität Wien (TU Wien). He is currently a Senior Scientist with TU Wien, with his
research focusing on the performance aspects
of mobile cellular technologies. He is currently
examining the feasibility of using crowdsourcing
to conduct performance measurements on 4G and
5G mobile networks. His research aims to establish
a common framework for evaluating the performance of mobile networks, guaranteeing reliable and fair connectivity for
end-users.
\end{IEEEbiography}

\clearpage
\newpage
\appendices

\section{Additional Reference and Frequency-Dependent Radiation Maps}
\label{app:additional_figures}

For completeness, this appendix collects the reference and frequency-dependent radiation maps that were removed from the main text to improve figure economy while preserving the full measurement context.

\begin{figure}[!h]
    \centering

    \begin{minipage}[b]{0.4\textwidth}
        \centering
        \includegraphics[width=\linewidth]{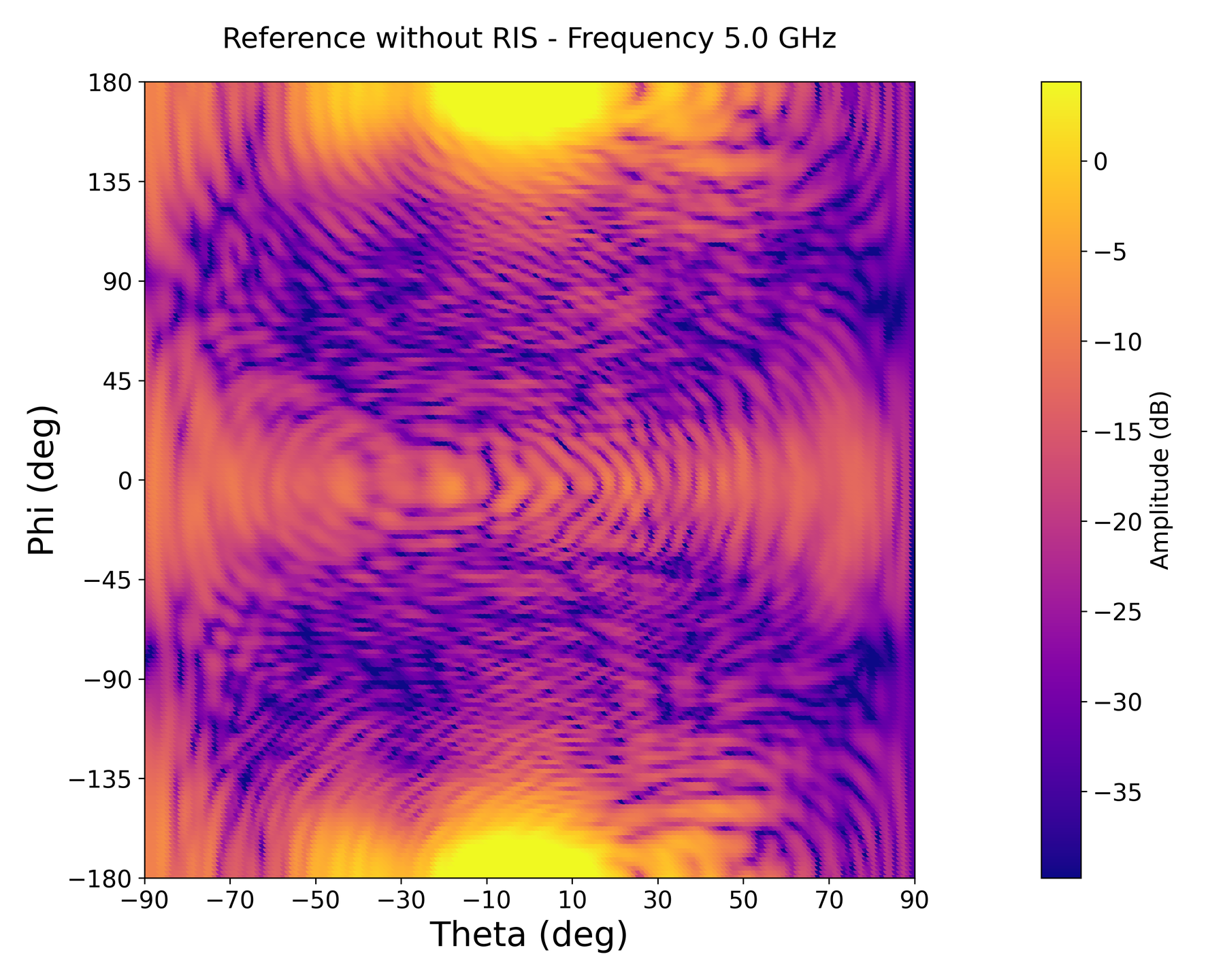}
        \vspace{1mm}
        \centerline{(a)}
    \end{minipage}
    \hfill
    \begin{minipage}[b]{0.4\textwidth}
        \centering
        \includegraphics[width=\linewidth]{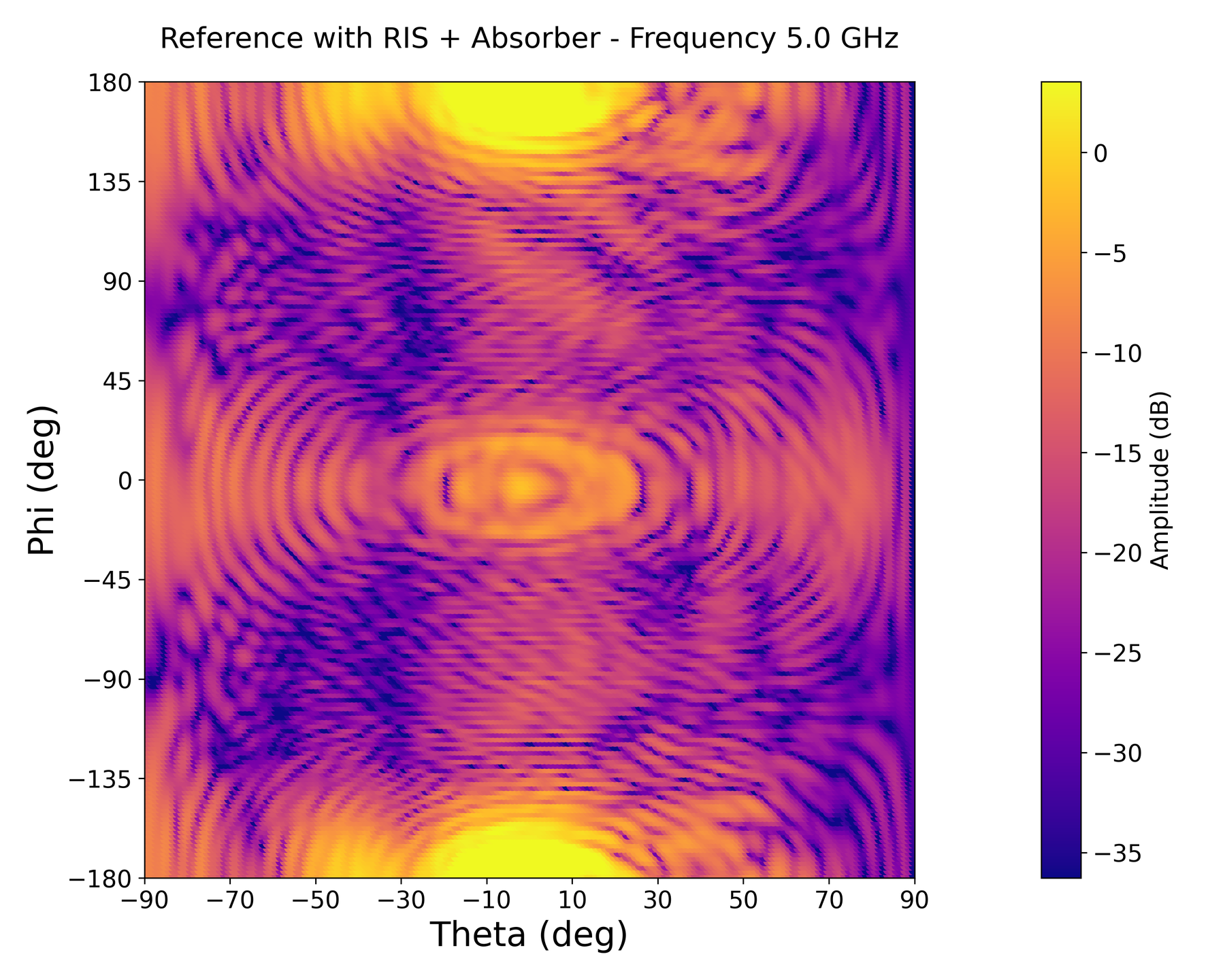}
        \vspace{1mm}
        \centerline{(b)}
    \end{minipage}

    \caption{Additional reference measurements moved from the main text:
    (a) reference without RIS, with electronics installed;
    (b) reference with RIS and added sheet absorber.}
    \label{fig:appendix_reference}
\end{figure}

\begin{figure}[!t]
    \centering

    \begin{minipage}[b]{0.35\textwidth}
        \centering
        \includegraphics[width=\linewidth]{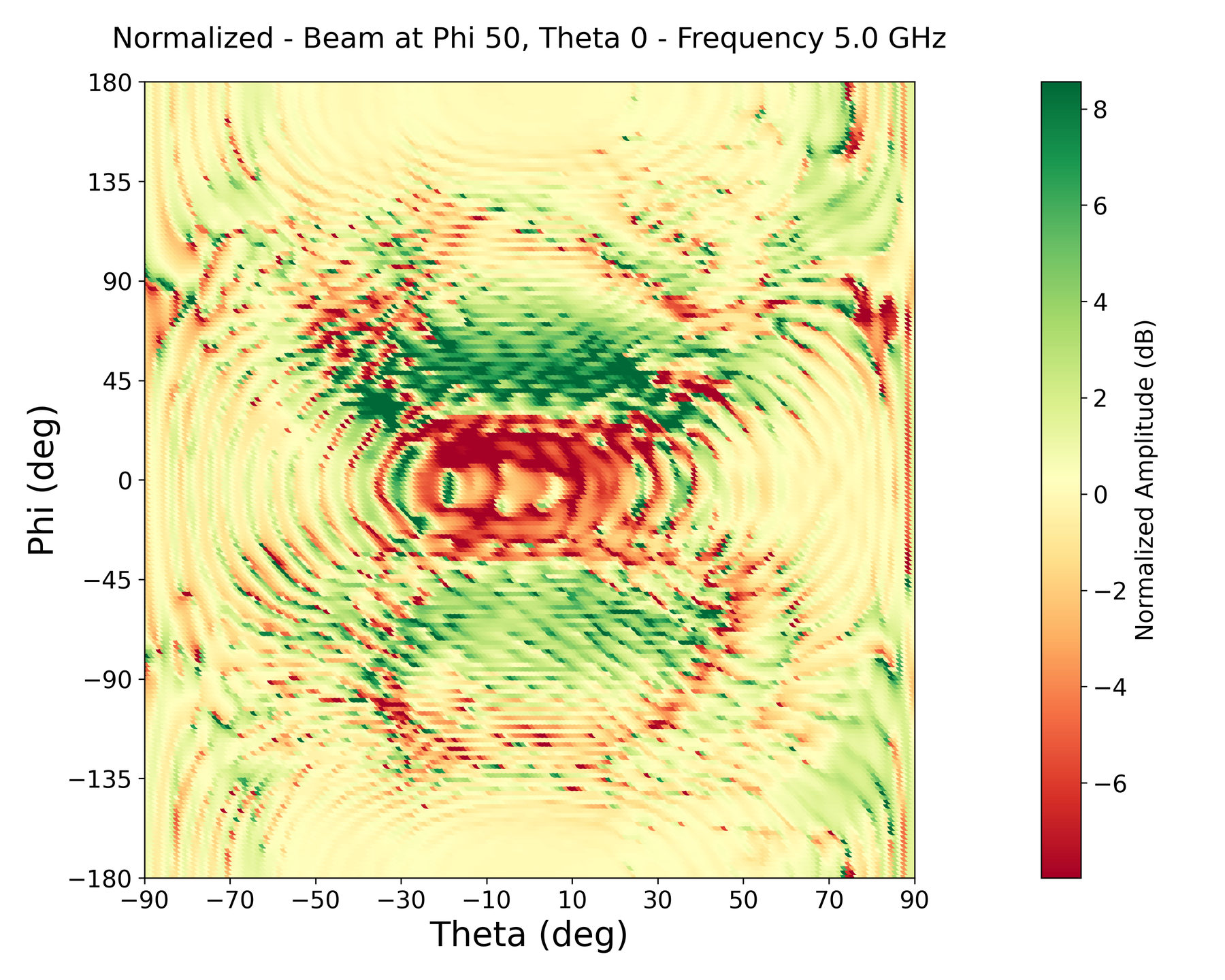}
        \vspace{1mm}
        \centerline{(a)}
    \end{minipage}
    \hfill
    \begin{minipage}[b]{0.35\textwidth}
        \centering
        \includegraphics[width=\linewidth]{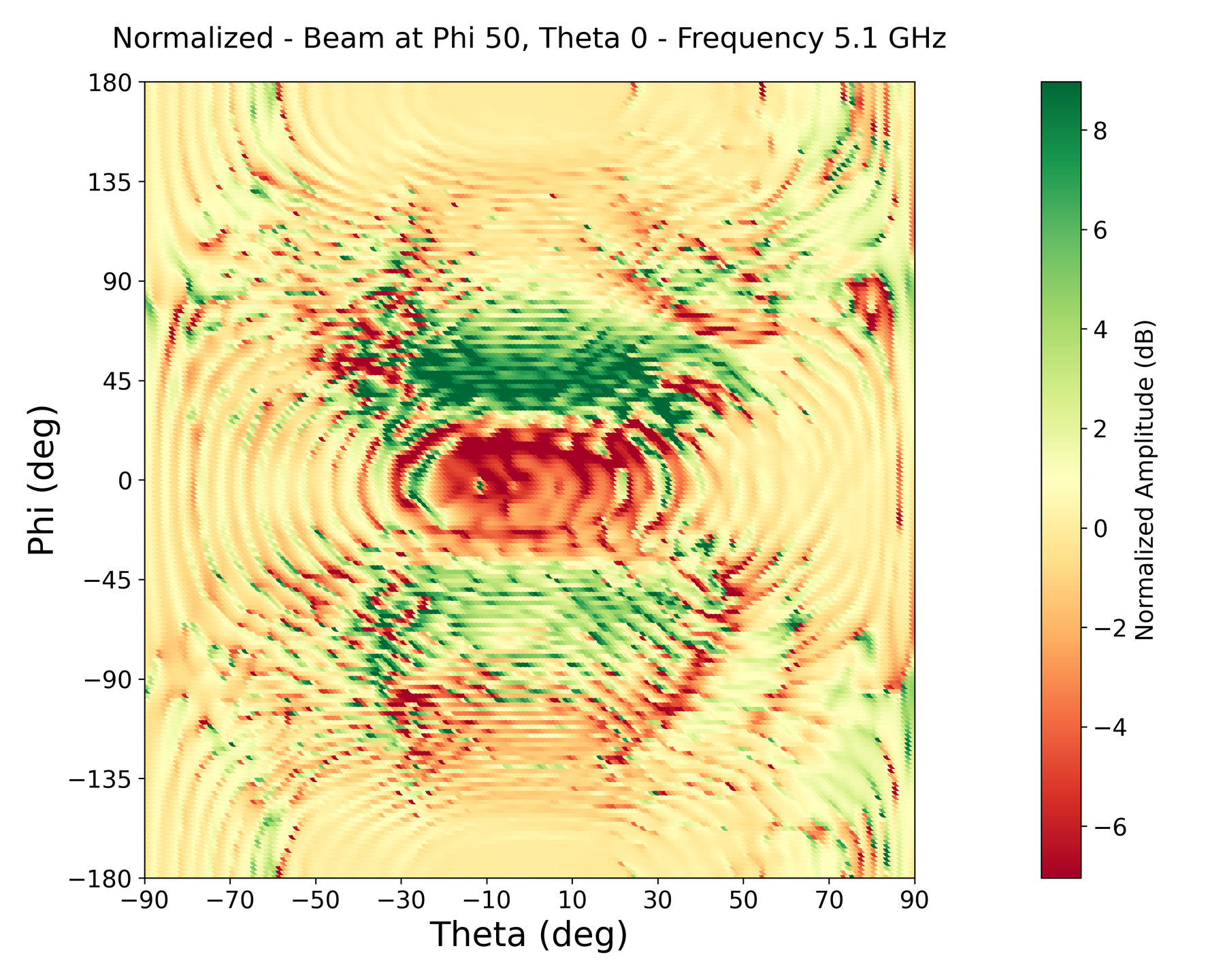}
        \vspace{1mm}
        \centerline{(b)}
    \end{minipage}

    \vspace{2mm}

    \begin{minipage}[b]{0.35\textwidth}
        \centering
        \includegraphics[width=\linewidth]{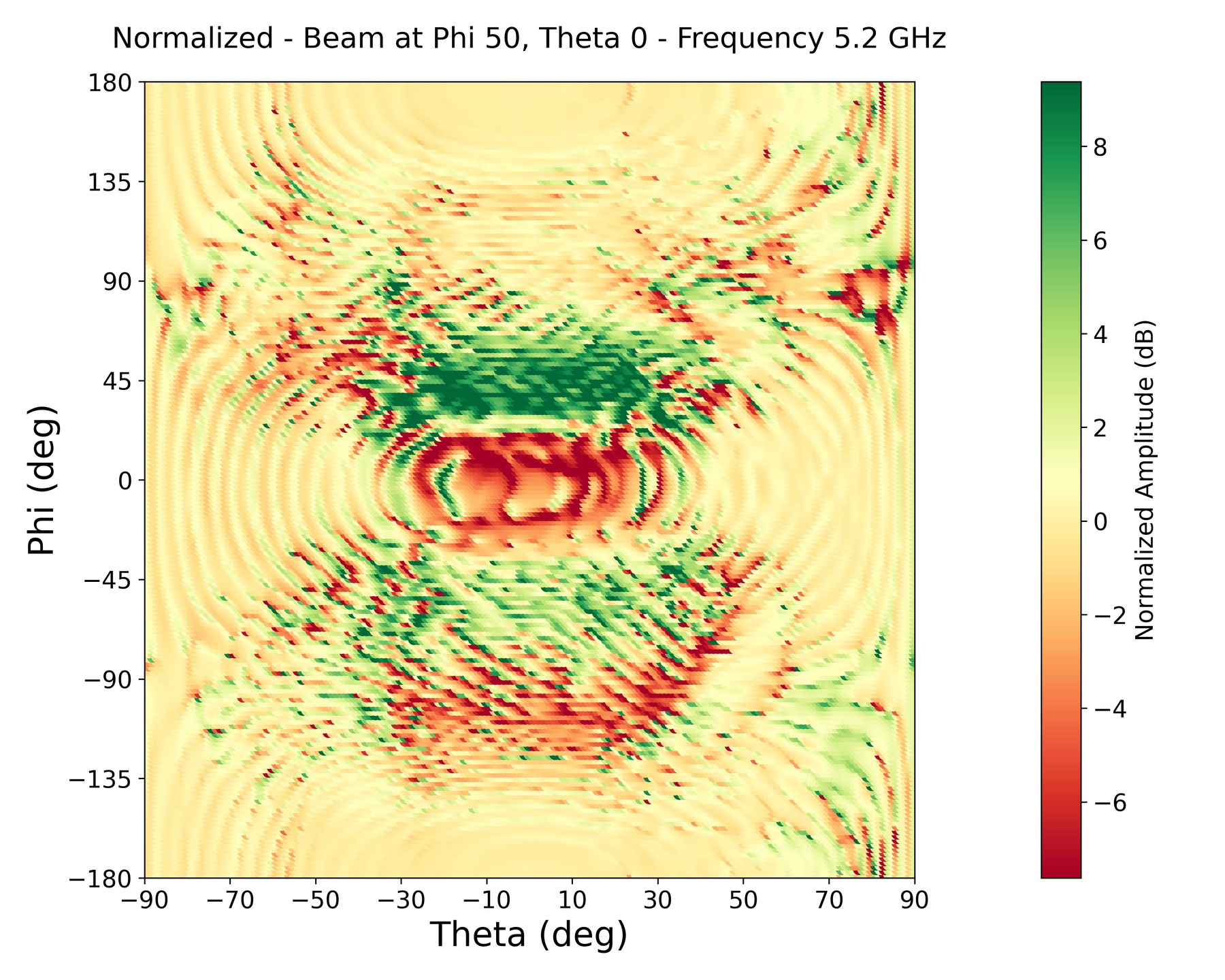}
        \vspace{1mm}
        \centerline{(c)}
    \end{minipage}
    \hfill
    \begin{minipage}[b]{0.35\textwidth}
        \centering
        \includegraphics[width=\linewidth]{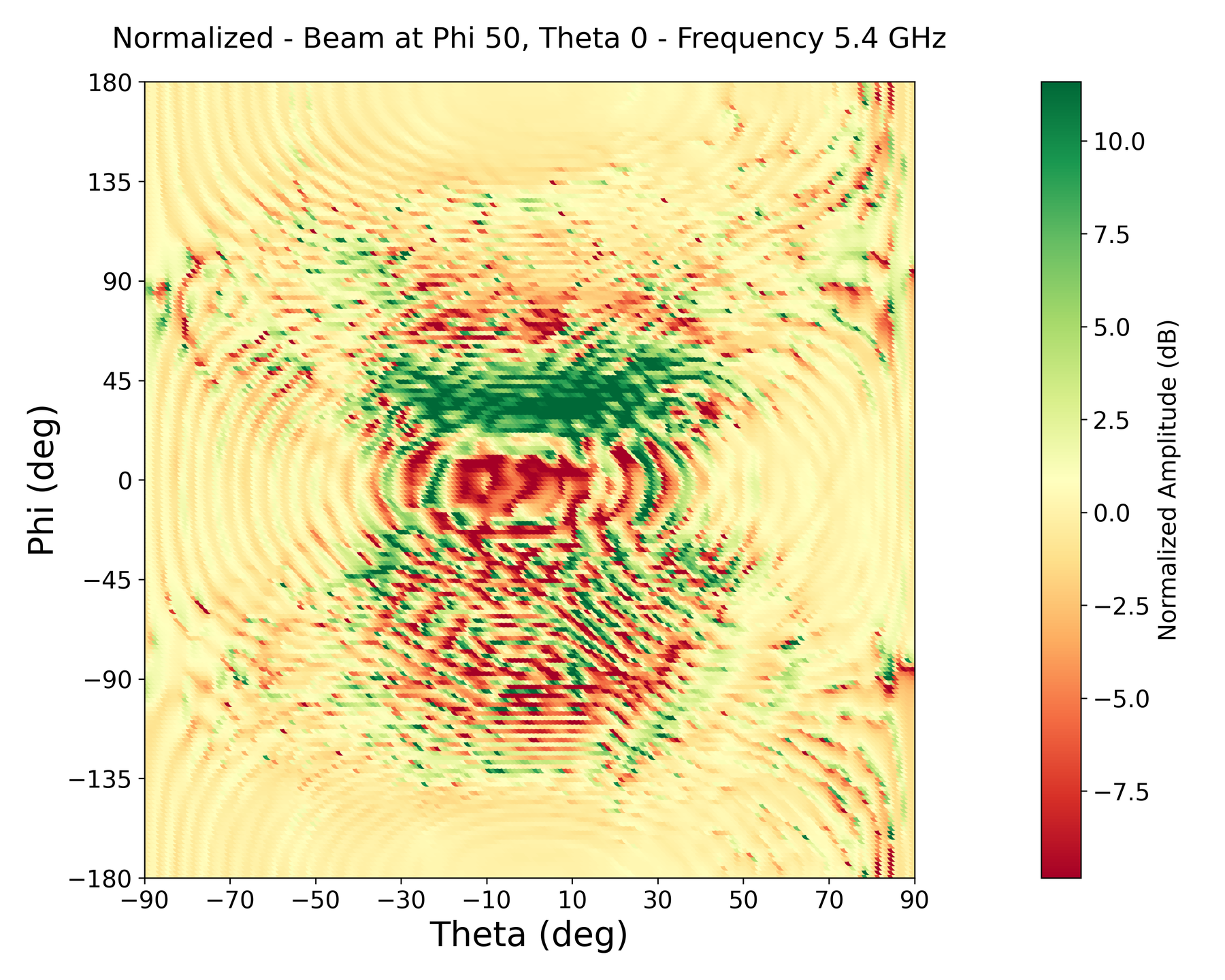}
        \vspace{1mm}
        \centerline{(d)}
    \end{minipage}

    \caption{Additional frequency-dependent radiation maps moved from the main text for the beam at $\phi=50^\circ$ and $\theta=0^\circ$:
    (a) 5.0~GHz;
    (b) 5.1~GHz;
    (c) 5.2~GHz;
    (d) 5.4~GHz.}
    \label{fig:appendix_frequency}
\end{figure}

\vspace{11pt}

\end{document}